\pdfoutput=1
\documentclass[fleqn,usenatbib]{mnras}

\usepackage[T1]{fontenc}

\DeclareRobustCommand{\VAN}[3]{#2}
\let\VANthebibliography\thebibliography
\def\thebibliography{\DeclareRobustCommand{\VAN}[3]{##3}\VANthebibliography}

\usepackage{graphicx}	% Including figure files
\usepackage{amsmath}	% Advanced maths commands
\usepackage{amssymb}	% Extra maths symbols
\usepackage{xcolor}
\newcommand{\beq}{\begin{equation}}
\newcommand{\eeq}{\end{equation}}
\newcommand{\Mdot}{\dot{M}}
\newcommand{\hii}{H~{\sc II}~}
\newcommand{\kms}{\mbox{km s$^{-1}$}}

\newcommand{\gcm}{\mbox{g cm$^{-3}$}}

\newcommand{\Mo}{\mbox{M$_{\odot}$}}
\newcommand{\Moy}{\mbox{M$_{\odot}$ yr$^{-1}$}}

\title[Planetary Nebula Evolution]{Planetary Nebula Evolution for Single Stellar Models. \\
The Formation of Neutral Spikes}

\author[Garc\'{\i}a-Segura et al.]{
Guillermo Garc\'{\i}a-Segura$^{1,2}$,\thanks{E-mail: ggs@astrosen.unam.mx (GGS)}
Arturo Manchado$^{3,4,5}$,
Jes\'us A. Toal\'a$^{6}$, 
Mart\'{\i}n A. Guerrero$^{2}$, 
\newauthor 
Alberto J. Castro-Tirado$^{2,7}$
\\
$^{1}$Instituto de Astronom\'{\i}a, Universidad Nacional Aut\'onoma
de M\'exico (UNAM), Apdo. Postal 877, 22800 Ensenada, B.C., Mexico \\
$^{2}$Instituto de Astrof\'{\i}sica de Andaluc\'{\i}a (IAA-CSIC), Glorieta de la Astronom\'{\i}a s/n, E-18008, Granada, Spain\\
$^{3}$ Instituto de Astrof\'{\i}sica de Canarias, v\'{\i}a L\'actea S/N, E-38200 La Laguna, Tenerife, Spain\\
$^{4}$ Departmento de Astrof\'{\i}sica, Universidad de La Laguna (ULL), E-38206 La Laguna, Tenerife, Spain \\
$^{5}$ Consejo Superior de Investigaciones Cient\'{\i}ficas, Spain\\
$^{6}$ Instituto de Radioastronom\'{\i}a y Astrof\'{\i}sica, UNAM Campus Morelia, Apartado postal 3-72, 58090 Morelia, Michoacan, Mexico \\
$^{7}$ Departamento de Ingenier\'{\i}a de Sistemas y Autom\'atica, Unidad Asociada al CSIC por 
el IAA, Escuela de Ingenier\'{\i}as Industriales,\\ 
Universidad de M\'alaga, C. Dr. Ortiz Ramos s/n, 
29071 M\'alaga, Spain}

\date{Accepted XXX. Received YYY; in original form ZZZ}

\pubyear{2022}

\begin{document}
\label{firstpage}
\pagerange{\pageref{firstpage}--\pageref{lastpage}}
\maketitle

\begin{abstract}
Two-dimensional hydrodynamical simulations are presented from the formation up to the late evolution of planetary nebula, for 6 different stellar models from 1 to 5 \Mo. 
The hydrodynamical models use stellar evolution calculations as inner boundary conditions
and updated values for the number of ionizing photons.
Special emphasis is placed on the formation of neutral spikes, as recently observed by the James Webb Space Telescope. The results indicate that neutral spikes can be detected either at the formation of planetary nebulae or in their decline. In the first case, the temporal window decreases with the mass of the model, ranging from 3,000 years in the 1 \Mo \, case to 0 for 5 \Mo. In the second case, only the 1.5, 2.0, and 2.5 \Mo \, cases allow us to detect the neutral spikes for most of the remaining time.

\end{abstract}


\section{Introduction}

Planetary nebulae (PNe) are descendent objects from the asymptotic giant branch (AGB) phase. 
Once the outer layers of the AGBs are removed (either by stellar mass loss or in a more violent
event such as common envelope evolution), the hot cores of the stars are exposed with high temperatures,
allowing the photoionization of their circumstellar media, which, at the same time, are being swept up
by the stellar winds, forming the classical shapes of ring nebulae (Kwok, Purton \& Fitzgerald 1978).
In these processes, the ionization fronts travel outwards to the point where the circumstellar gas
is totally photoionized, creating spectacular PNe. Later on, the PN central stars enter the
cooling tracks on their way to white dwarfs, reducing the number of ionizing photons. At this point,
the ionizing fronts retreat and the PNe are expected to recombine.
Since PNe are subject to dynamical instabilities, the way ionizing fronts produce the initial
breakout or their retreat depends on the opacity in the line of sight, which is very sensitive to the
existence of clumps, filaments, or any kind of over densities as seen in high-resolution images
(Manchado et al. 2015; Wesson et al. 2024).

Infrared observations of PNe have shown the existence of molecular hydrogen emission
in many objects (Kastner et al. 1994; Guerrero et al. 2000; Hora et al. 2006; Akras et al. 2020), probing that in many cases the ionization front is trapped in the nebulae. Recently, the James Webb Space Telescope (JWST) has observed two planetary nebulae,
NGC 3132 (the southern ring nebula, De Marco et al. 2022)  and NGC 6720 (the ring nebula, Wesson et al. 2024), where both showed multiple neutral spikes detected in molecular hydrogen filters. 
Motivated by these observations, we conducted this study in order to show in which range of solar masses and
in which part of their evolution these molecular features could form. This study is a two-dimensional new version
of the work done by Villaver et al. (2002), with a correction in the number of ionizing photons as 
given by Toal\'a \& Arthur (2014). The main difference
from the work by Toal\'a \& Arthur (2014), beside differences in   geometry and resolution, is
that we follow the evolution to longer times and focus our study in the neutral gas instead of the hot coronal
gas.

The point of this paper is to show that the models presented illustrate that clump and spike formation in simply structured PNe is a very plausible outcome of thin-shell instabilities (under all conditions considered), and that clumps and spikes are likely to be common at times in phases A and C of PN evolution (explained below) for every initial mass of the central star. These models must be understood as qualitative rather than quantitative results, and give a first approximate look at how spikes can form in PNe. In the future, more sophisticated and precise models are needed for fine-tuning.

This article is structured as follows: the numerical scheme and physical approximations, as well as the inputs for our calculations, are described in \S~2.  The results of the numerical simulations are presented in \S~3.   Finally, we discuss the numerical results in \S~4 and provide the main conclusions in the last section.

\section{Numerical methods and physical assumptions}

The numerical simulations have been performed using the magnetohydrodynamic code
ZEUS-3D (Version 3.4), developed by M. L. Norman and the Laboratory for
Computational Astrophysics. It is a fully explicit Eulerian code of finite differences that descends from the code described in Stone \& Norman (1992) and Clarke (1996). A simple approximation is used to derive the location of the ionization
front in this study (Garc\'{\i}a-Segura \& Franco 1996),
assuming that the ionization equilibrium applies at all times. This approximation is very useful
for long runs, since more complicated schemes could slow down the computation considerably,
since each run took around 40 days.
The models include the Raymond \& Smith (1977) cooling curve above $10^6$ K.
For temperatures below $10^6$ K, the cooling curve given by  MacDonald \& Bailey (1981) is
applied. We have optimized the cooling curve for average planetary nebula abundances, as shown in 
Table 1 of Toal\'a \& Arthur (2014), with, for example, N = 8.2 and O = 8.6 (lg X + 12). 
These values remain constant throughout the simulation.

\begin{figure}
\includegraphics[width=\linewidth]{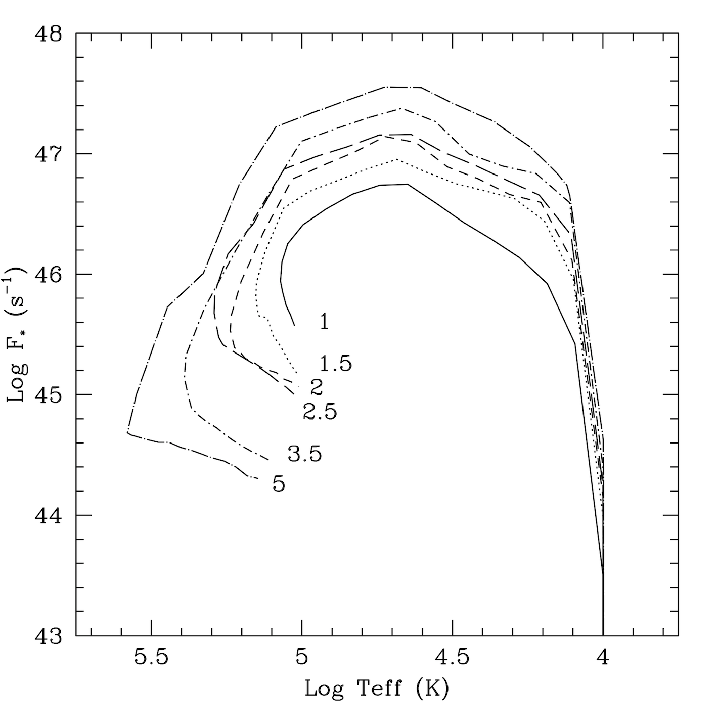}
\caption{Total number of ionizing photons per unit time. Labels correspond to ZAMS masses.}
\label{f1}
\end{figure}

\begin{figure}
\includegraphics[width=\linewidth]{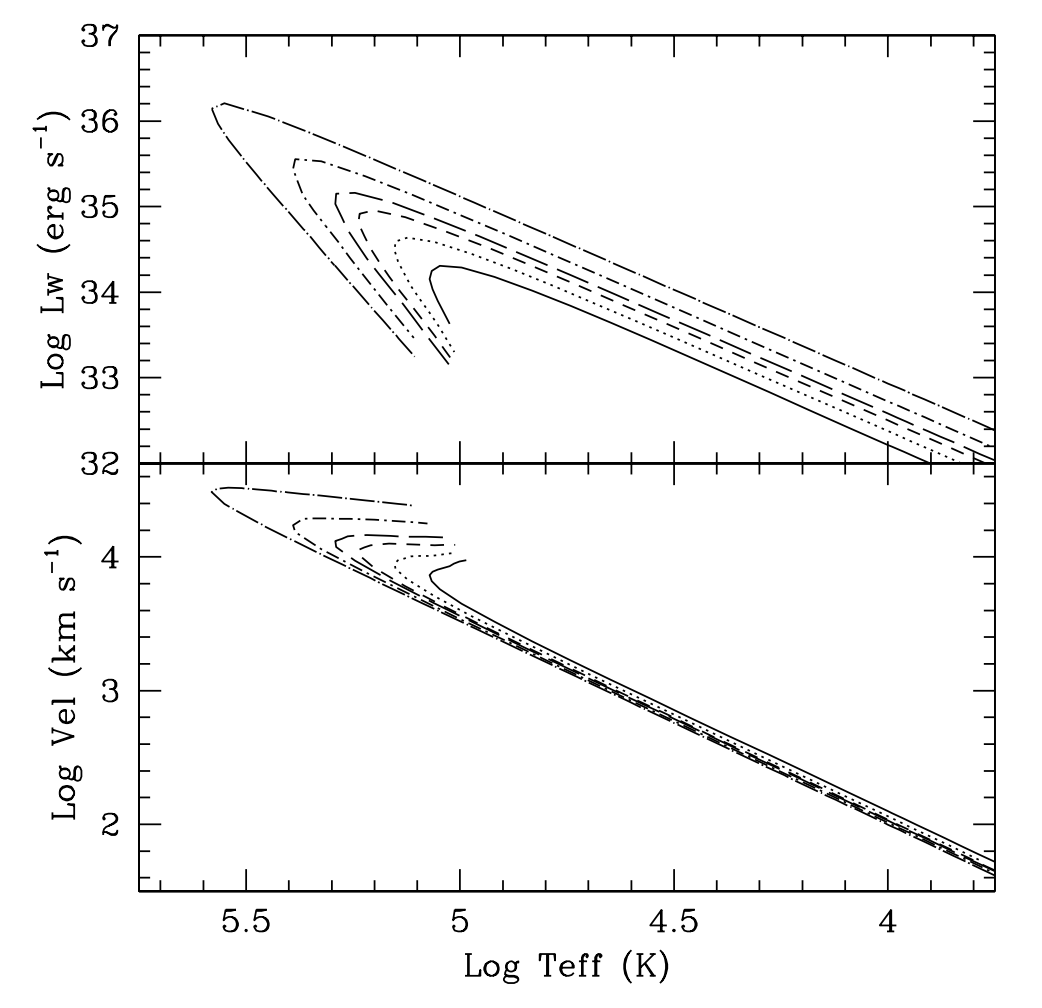}
\caption{Wind kinetic energy (mechanical luminosity $\rm L_w$ ) (top) and wind velocity ($\rm Vel$) (bottom) . Line styles similar to Figure 1.}
\label{f2}
\end{figure}

\begin{figure}
\includegraphics[width=\linewidth]{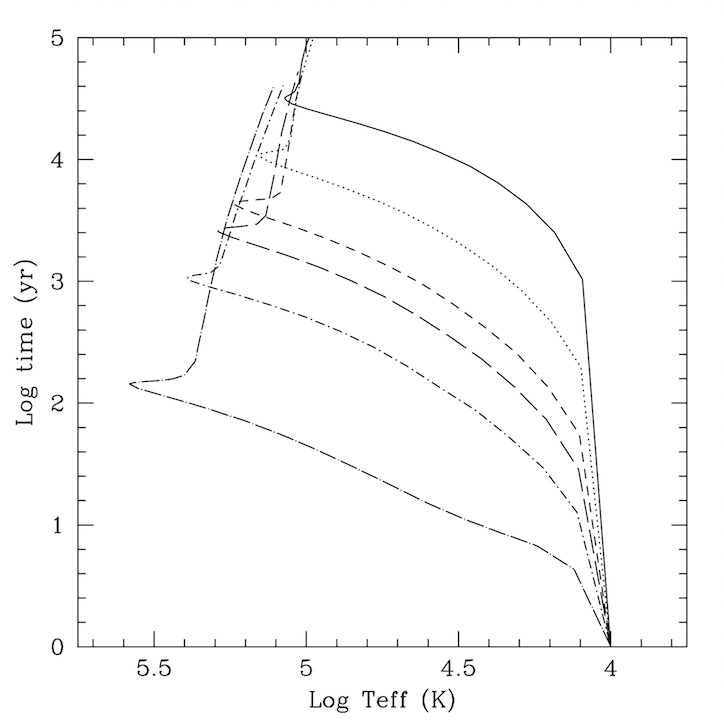}
\caption{Evolution of the effective temperature from the begining of photoionization. Line styles similar to Figure 1.}
\label{f3}
\end{figure}

The numerical calculations use a 2-dimensional computational grid in spherical coordinates with
$500 \times 200$ equidistant zones in $r$ and $\phi$, respectively, with an angular extent of $22.5^{\circ}$, and an
initial radial extent of 0.001 pc, giving an initial resolution of $2 \times 10^{-6}$ pc per radial zone.
A self-expanding grid technique has been employed in order to allow the radial coordinate
to grow by several orders of magnitude. The tracking of the grid is focused on the largest velocity in
the middle of the radial coordinate, in order to have a large fraction
of the grid ($\sim$ 50 \%) to develop the expected neutral spikes outside of the swept-up shell.

The AGB complete evolution for each stellar model is run previously in a 1-D spherically
symmetric grid, with 5,000 radial zones that cover 2.5 pc, similar (but with higher resolution) 
to the work by Villaver et al. (2002).  
The last outputs from these 1-D simulations are used as outer boundary conditions in the 2-D runs, i.e., 
the external AGB wind, 
interpolating linearly the values for the variables at the outer boundary while the 2-D grids expand during
the evolution in the former medium. 

External to the AGB wind, the interstellar medium density is set to 1 ${\rm cm}^{-3}$, 
which corresponds to moderate densities inside the Galactic spiral arms (Villaver et al. 2012).
Values of 0.1 or 0.01 ${\rm cm}^{-3}$ are probably more realistic for many PNs outside spiral arms, but this choice
gives too big sizes for the AGB wind stagnation point, and initial grids larger than 2.5 pc are needed. However, this is not necessary for our study.

Injection of mass and momentum into the computational grid representing the stellar wind is treated as an inner boundary condition, covering the two innermost
radial zones. The history of the winds are taken from Villaver et al. (2002), based on the stellar
evolution models for zero-age main sequence (ZAMS) masses 1, 1.5, 2.0, 2.5, 3.5 and 5 \Mo \, from Vassiliadis \& Wood (1994). The number of ionizing photons has been corrected according to Toal\'a \& Arthur (2014). Finally, periodic boundary conditions are used for the inner and outer boundary
conditions in $\phi$, which means that if some parcel of gas leaves the grid at the outer $\phi$ boundary, 
it will return to the inner $\phi$ boundary or the other way around. 
Figures 1 and 2 show the total number of ionizing photons per unit time,
the amount of kinetic energy per unit time (mechanical luminosity $\rm L_w$) and the wind velocity
injected into the grids as a function of the effective temperature. 
These figures are also shown as a function of time in Villaver et al. (2002) and in Toala \& Arthur (2014).
Figure 3 shows the evolution of the effective temperature as a function of time after photoionization starts.

\section{Results}

The six calculations begin when the stars leave the AGB phase, in which the velocity of the
wind becomes larger than 15 \kms . Thus, after a period of time, also called transition time, 
where the post-AGB wind starts
sweeping up the AGB wind,  photoionization begins when $\sim 10^4$ K is reached by the central star, defining
the beginning of the PN phase properly ($t_*=0 \,\, \, {\rm when} \, \,\, T_{eff}=10^4$ K). 
This transition time is based on the recipe for the mass loss
rate applied in the stellar evolution calculations for isolated stars. In the case of common
envelope evolution, this transition time could be quite different but is still unknown since it will
depend on the efficiency of the envelope ejection for each binary case. 

As a common scenario for all stellar models, there is an initial phase, {\bf Phase A}, where the ionization front makes its way to achieve photo-ionization of the entire nebula.
Some of the gas in this phase is still neutral.
After that, {\bf Phase B} is the phase in which photoionization is completed, where neutral gases do not coexist in the nebula. This is probably resolution dependent and will be discussed later on
 \S~4. This phase B is also called a density-bounded PN. Afterward, if the number of ionizing photons decreases considerably towards the cooling track, {\bf Phase C} begins where recombination succeeds and the ionization front retreats, giving rise to neutral gas again. 
 This phase C is also called an ionizing bounded PN. 
 
In the next subsections, the different phases are shown for the six stellar models, displaying
a whole view of the total evolution for each case. 

\subsection{Explanation of the figures}

Each figure for phases A and C shows three different views of the gas density evolving in time. 
Clockwise in each screenshot, we first show the neutral gas density in the first zone of $22.5^{\circ}$.
Black means no neutral gas (i.e., photoionized gas). In the second zone of $22.5^{\circ}$ we show
the photoionized gas density. Similarly, black means no ionized gas (i.e., neutral). The remaining $45^{\circ}$
shows the total gas density repeated twice. In this manner, we can observe, in the same figure, which part is neutral, which part is ionized, and a whole view of the total gas density in the third part. The labels in the upper right part of each screenshot correspond to the physical time $t_{\rm AGB}$ of the evolution after AGB in years. 
In parentheses, the time $t_*$, whose origin is $t_*=0$ when $T_{\rm eff}=10,000 $ K. 
Figures with phase B only show photoionized gas density, since there is no neutral gas.
We now describe in detail each of the stellar models. 
The models are summarized in Table 1, where $F_{\star} $ is the number of ionizing photons.

\begin{table}
\caption{Representative numbers for models. In parenthesis the mass of the star at the PN phase.}
\begin{tabular}{llrrrr}
\hline
 $ Model $ & $\,$  &  $t_{\rm AGB}$ & $t_{*}$ & $T_{\rm eff}$ & Log $F_{\star} $  \\
 $\,$ & $\,$  &    yr  &           yr          &      K        &   $s^{-1}$ \\
\hline
 1.0 \Mo &  Phase A & 9,100 & -100  & 9,775  &  43.19\\
 (0.569 \Mo)  &  Phase B & 15,300 & 6,100 & 21,200 & 46.22   \\
 $\,$   &  Max. $F_{\star}$ & 23,109 & 14,009 & 44,463 & 46.75  \\
$\,$   &  Phase C & - & - & - & -  \\
$\,$   &  Max. $T_{\rm eff}$ & 41,153 & 32,054 & 117,760 & 45.94 \\
\hline
1.5 \Mo &  Phase A & 6,900 & -860  & 9,447  &  43.68   \\
 (0.597 \Mo)   &  Phase B & 8,900 & 1,140   & 30,831 & 46.75   \\
 $\,$    &  Max. $F_{\star}$ & 11,444 & 3,704 & 48,305 & 46.95 \\
$\,$    &  Phase C & 17,400 & 9,640 & 141,579 & 45.93 \\
$\,$    &  Max. $T_{\rm eff}$ & 18,719 & 10,978 & 142,332 & 45.81\\
\hline
2.0 \Mo &  Phase A & 9,400 & -1,160 & 9,416  &   43.85 \\
 (0.633 \Mo)   &  Phase B & 11,200 & 640  &  32,885  & 46.89 \\
 $\,$    &  Max. $F_{\star}$ & 11,750 & 1,310 & 52,966 & 47.14 \\
 $\,$   &  Phase C & 13,700 & 3,140 & 134,276 & 46.34 \\ 
 $\,$    &  Max. $T_{\rm eff}$ & 14,887 & 4,447 & 172,981 &  45.53 \\ 
\hline
2.5 \Mo &  Phase A & 9,800 & -1,220 & 9,512 & 43.93 \\
 (0.677 \Mo)   &  Phase B &  11,400 & 380 & 38,656 &  47.08   \\
 $\,$    &  Max. $F_{\star}$ & 11,506 & 526 & 43,351 & 47.16 \\
 $\,$   &  Phase C & 12,800 & 1,780 & 143,877 & 46.43 \\ 
 $\,$    &  Max. $T_{\rm eff}$ & 13,646 & 2,666 & 195,884 & 45.66 \\ 
\hline
3.5 \Mo & Phase A & 18,300 & -2,130 & 9,399 &  44.09 \\
(0.754 \Mo)   &  Phase B & 20,600 & 170 & 36,057 &  47.26 \\
$\,$    &  Max. $F_{\star}$ & 20,629 & 199 & 46,558 & 47.38 \\
$\,$   &  Phase C & 21,100 & 670 & 129,717 & 46.65 \\ 
$\,$    &  Max. $T_{\rm eff}$ & 21,497 & 1067 & 245,470 & 45.14 \\
\hline
5.0 \Mo & Phase A & 25,100 & -2,880 & 9,406 &  44.31 \\
(0.90 \Mo)    &  Phase B & 28,000 & 20 & 40,179 & 47.55 \\
$\,$    &  Max. $F_{\star}$ & 28,010.3 & 20.3 & 53,088 & 47.55 \\
$\,$   &  Phase C & 28,100 & 120 & 180,717 &  45.44 \\
$\,$    &  Max. $T_{\rm eff}$ & 28,131.7 & 141.7 & 381,065 &  44.68 \\
\hline

\end{tabular}

\end{table}

\subsection{The evolution of the 1  $\Mo $ model}

\begin{figure}
\includegraphics[width=\linewidth]{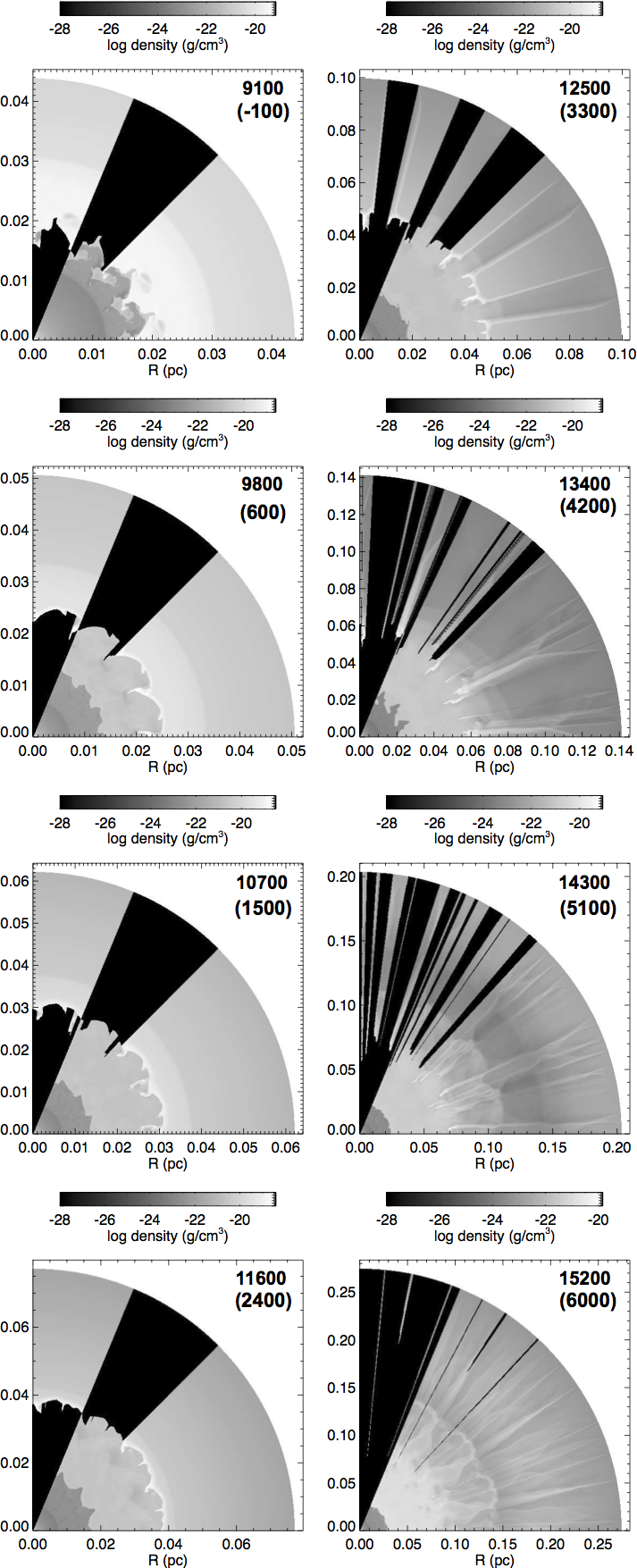}
\caption{Phase A for 1.0 \Mo. Each snapshot shows clockwise neutral gas density, ionized gas density and total gas density
as explained in subsection 3.1 }
\label{f4}
\end{figure}

\begin{figure*}
\centering
\includegraphics[width=\linewidth]{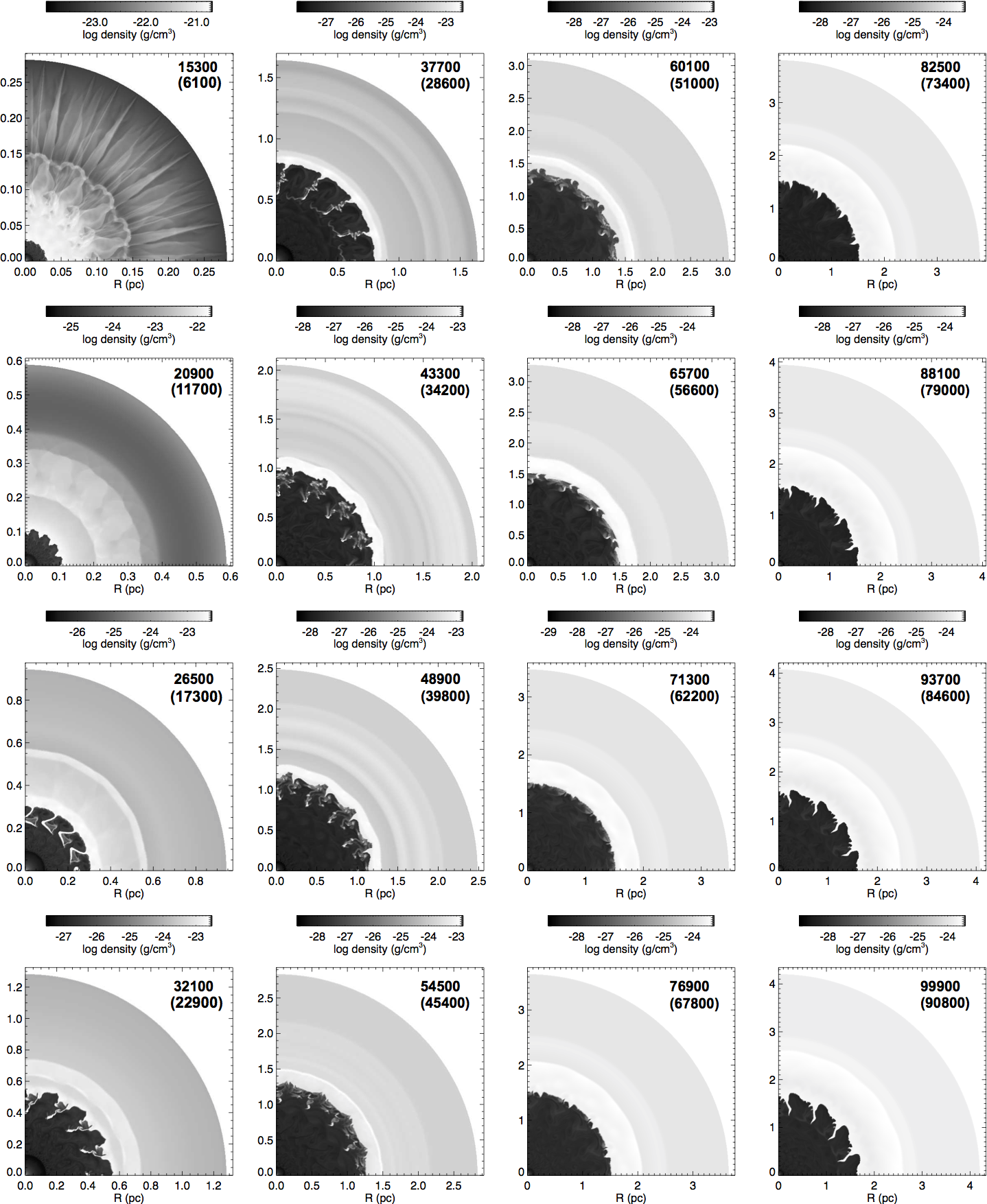}
\caption{Phase B for 1.0 \Mo}
\label{f5}
\end{figure*}

The evolution of the 1 $\Mo \, $ model is shown in Figures 4 and 5. Figure 4 shows phase A for this model, 
while Figure 5 shows phase B. This model does not have phase C. 
The first snapshot in Figure 4 is at 9100 years from the AGB departure, when the
star reaches 9,775 K and begins to photoionize. The young planetary nebula
at the beginning of phase A has an outer size of 0.03 pc,
an internal wind shock located at 0.012 pc, and a small hot gas zone whose
contact discontinuity reaches 0.02 pc, with temperatures of $10^5$ K
making it very radiative. This mode of dynamics is called momentum conserving (Treffers \& Chu 1982).

In the second snapshot of Figure 4, 700 years later,
an \hii region has been generated,
ranging from 0.014 pc to 0.0248 pc,
whose high pressure has created a new external shock located at 0.0255 pc (D-type front),
which in turn has compressed the hot gas region towards the center.
The internal shock has receded to 0.007 pc while the hot region
now has its contact discontinuity at 0.014 pc.
The hot gas still has temperatures of $10^5$ K since the wind velocity is relatively low ($\sim 10^2 \, \kms$).
Between the ionization front at 0.0248 pc and the new external shock at 0.0255 pc, a shell of swept-up
shocked neutral gas of high density
($\sim 10^{-19}$ \gcm) has been generated.

In the fifth snapshot of Figure 4 at 12,500 (3,300) years,
the ionization front has outgrown the grid size in the angular regions
where the nebula became optically thin, rapidly erasing all traces of the swept-up  neutral shell in those places, 
and is trapped
in the regions where the swept-up  neutral shell is denser. This leads
to an imbalance in the thermal pressure of the gas in the outer region
of the free AGB wind. This produces a lateral expansion of the ionized gas at
$10^4$ K, forming new lateral shocks, D-type fronts, that sweep
neutral material at $10^2$ K in the $\phi$ direction. 
The internal wind shock is located at 0.0075 pc and the contact discontinuity at 0.018 pc,
while the remaining swept neutral shell is located at 0.048 pc.

In the sixth snapshot (Figure 4) at 13,400 (4,200) years,
the neutral zones have been compressed into spikes.
The internal shock has shifted to 0.01 pc and the contact discontinuity to 0.02 pc. 
The clumps at the head of the spikes are located
between 0.05 and 0.06 pc. The \hii region extends from 0.02 pc to the end of the grid.
In this phase, the cometary heads are surrounded by photoionized gas
at $10^4$ K and not by hot coronal gas.

In the last two snapshots of Figure 4 at 14,300 (5,100) and 15,200 (6,000) years,
the spikes are disappearing since the number of ionizing photons is increasing considerably
by 3 orders of magnitude (Figure 1) and the neutral material becomes optically thin
to ionizing radiation. This is the end of phase A.

Figure 5 shows phase B for the 1 $\Mo \, $ model.
The first snapshot shows the beginning of phase B where the entire grid has been photoionized. 
The terminal wind shock is located
at 0.01 pc and the contact discontinuity is located at 0.03 pc. From here, a large HII region extends to the end of the grid with much substructure.

From the second snapshot at 20,900 (11,700) years, we find the classic planetary nebula structure, with an inner shock located at 0.035 pc, a swept-up shell at 0.1 pc, and a detached halo ranging from 0.21 to 0.34 pc.
The halo still retains some substructure.
From now on, the expansion is governed by wind energy, which reaches its maximum at 41,154 (32,054) years, just after the fifth
snapshot at 37,700 (28,600) years. After this, the wind begins to cease,
so the thermal pressure of the hot bubble slowly falls, which makes
the swept-up shell located at 1 pc start to thicken from the sixth snapshot at
43,300 (34,200) years. The internal shock is located at 0.1 pc. 

In the last snapshot of Figure 5 at 99,900 (90,800) years,
we observe a very faint and diffuse nebula, with a density only slightly
above the ISM, with sizes greater than 2 pc, which should be very difficult to observe with respect to the galactic background.

As we have seen in this model, there is only a small fraction of the nebula lifetime ($\sim$ 3,000 yr) where
neutral spikes could be observed at the end of phase A. After that, no neutral structures are formed in the
rest of the evolution. 

\subsection{The evolution of the  1.5 $\Mo $ model}

\begin{figure}
\includegraphics[width=\linewidth]{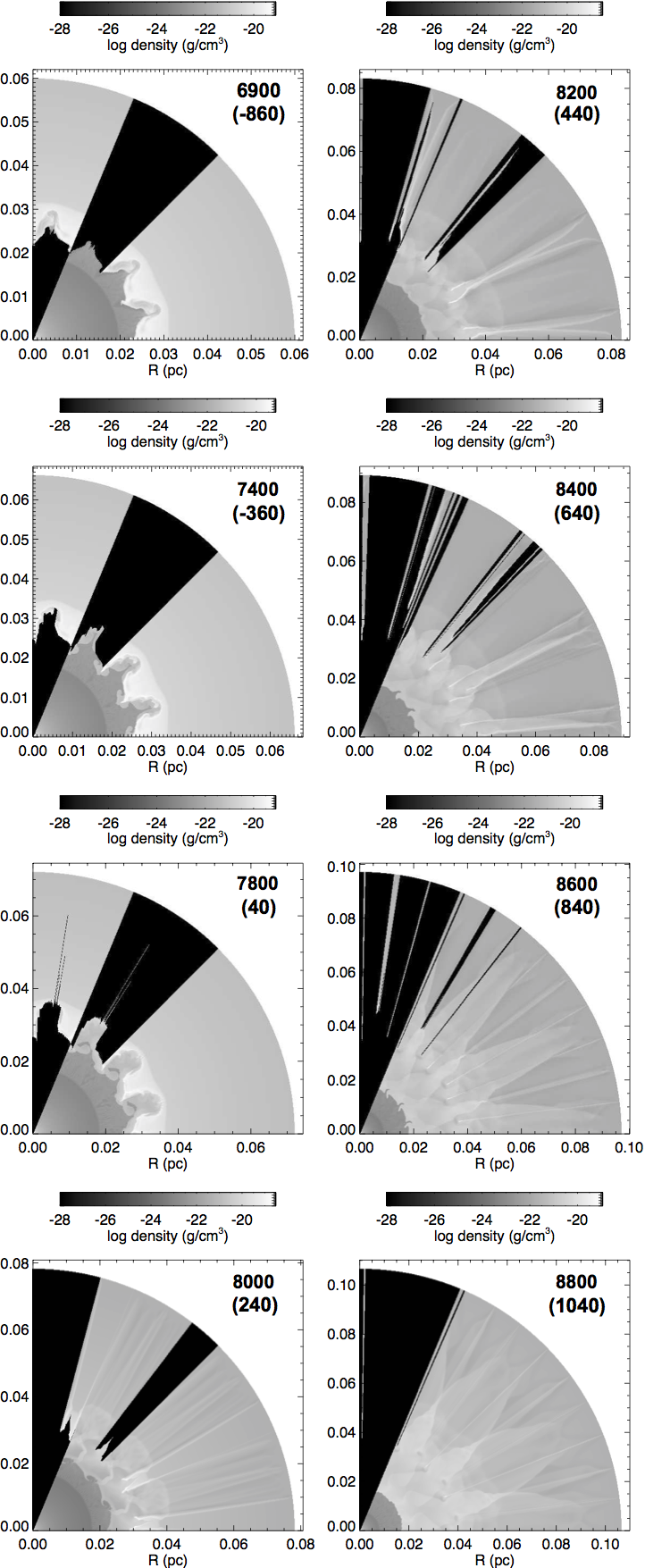}
\caption{Phase A for 1.5 \Mo}
\label{f6}
\end{figure}

\begin{figure}
\includegraphics[width=\linewidth]{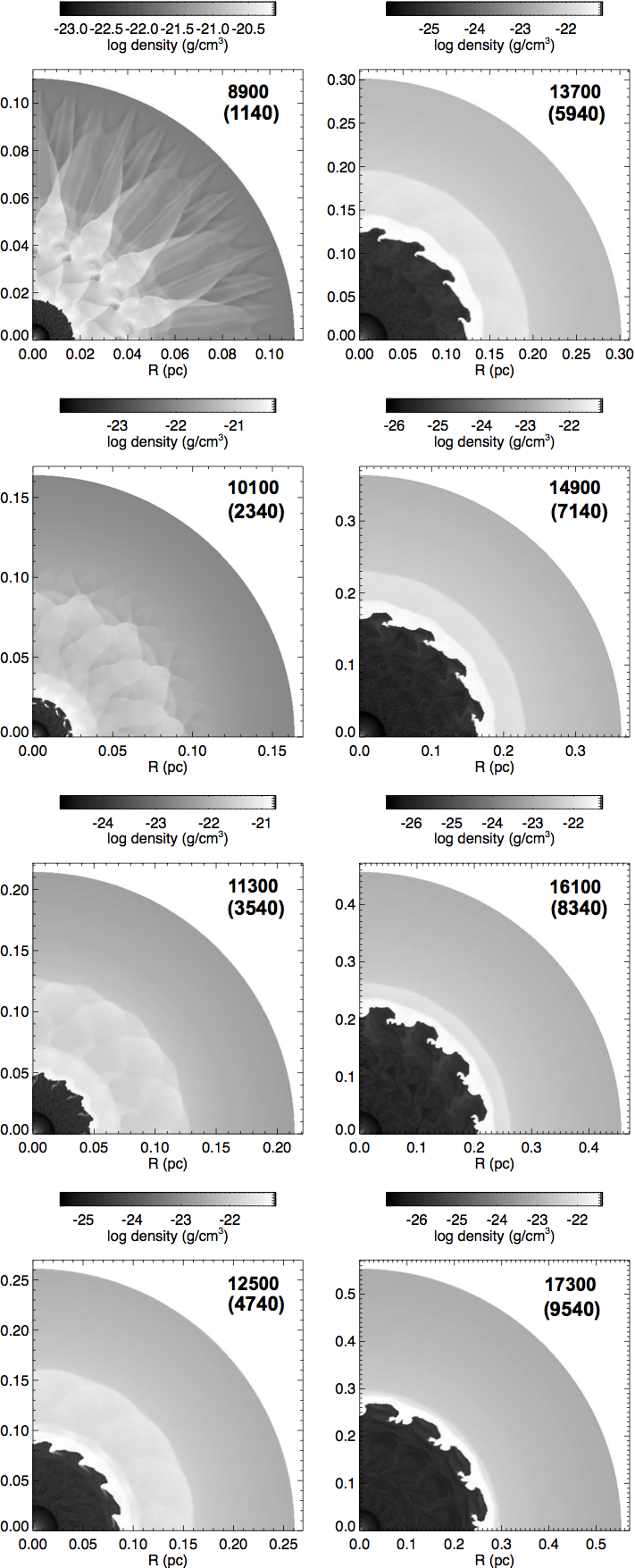}
\caption{Phase B for 1.5 \Mo}
\label{f7}
\end{figure}

\begin{figure}
\includegraphics[width=\linewidth]{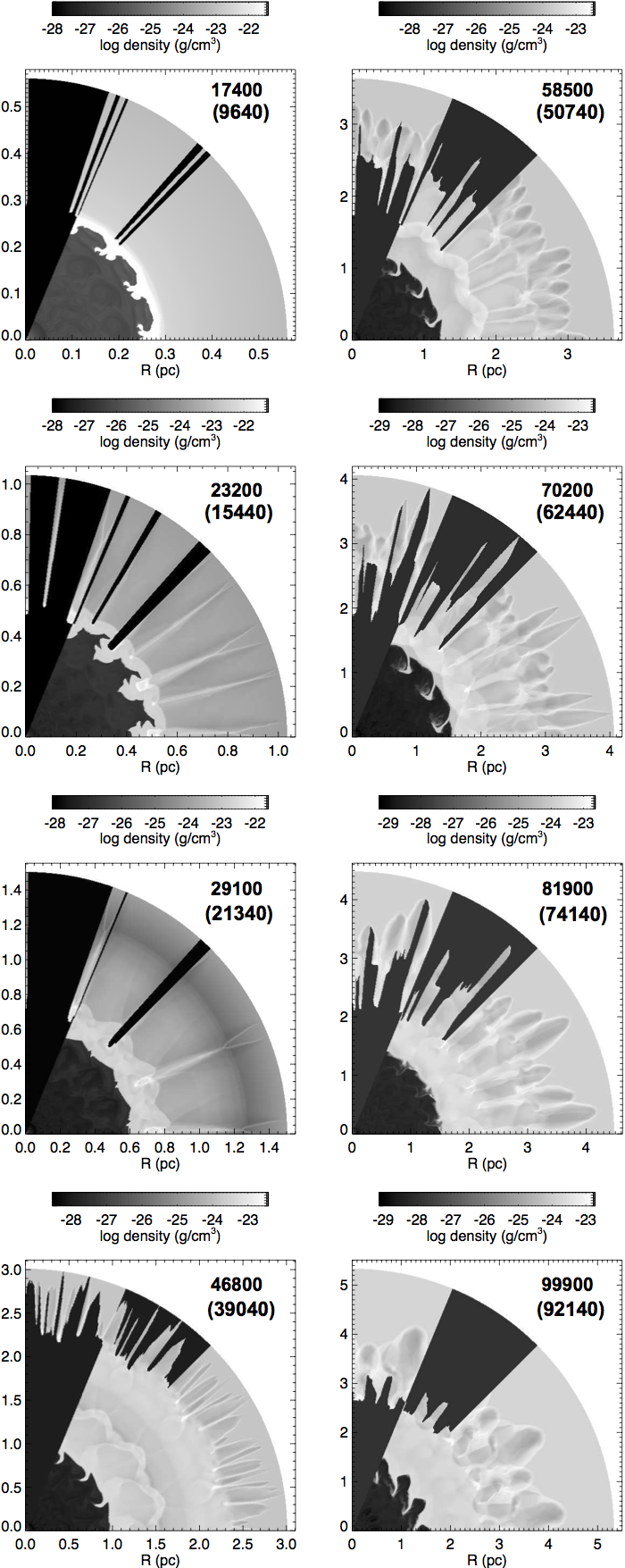}
\caption{Phase C for 1.5 \Mo}
\label{f8}
\end{figure}

The 1.5 $\Mo \,$  model is shown in Figures 6 (phase A), 7 (phase B), and
8 (phase C). Starting from the first snapshot in Figure 6,
we see that the protoplanetary nebula has a size of 0.031 pc in its outer shock and
an inner shock located at 0.019 pc. In the third snapshot at 7,800 (40) years,
the ionization front begins to advance beyond the outer shock.
In the fourth snapshot at 8,000 (240) years, the first neutral spikes are
already visible, which last for about 800 years until the end of phase A at 8,800 (1,040) years.

Moving on to phase B in Figure 7, we see in the first snapshot at 8,900 (1,140) years
a very extensive \hii region with a lot of substructure due to the thermal expansion of the old neutral spikes.
This substructure is still visible 2,400 years later in the third snapshot.
From the fourth snapshot at 12,500 (4,740) years to the eighth at 17,300 (9,540) years, we observe
a planetary nebula with an attached halo that is finally swept away in the last
snapshot. Here, the outer shock is located at 0.285 pc and the inner shock at 0.045 pc.

In phase C in Figure 8, the recombination starts in part of the nebula,
beginning to form neutral spikes that are visible already in the second snapshot
at 23,200 (15,440) years. According to Table 1, the maximum effective temperature is reached at 18,719 (10,978) years, which marks the decrease in the kinetic energy of the wind, which is why in the second snapshot it can be seen that the swept-up shell has begun to expand with a notable drop in its density, due to the fact that the thermal pressure of the bubble is rapidly falling, which is the pressure that drives the expansion of the planetary nebula. In the fourth snapshot at 46,800 (39,040) years
the thermal pressure of the bubble has dropped so much that the swept-up shell has practically disappeared, and the neutral clumps have also disappeared due to the thermal expansion, which have become
optically thin, marking the end of the neutral spikes and giving rise to
another more external structure because the ionization front has contracted, thus forming new neutral structures at the periphery of the computational grid.
In the fifth snapshot at 58,500 (50,740) years, neutral spikes are forming again because
the ionizing photons have decreased considerably and the conditions for the ionization front to be trapped in clumps are again given.
Thus, in conclusion, we could say that the existence of molecular spikes in
this 1.5 Mo model occupies a large part of phase C.

\subsection{The evolution of the 2 $\Mo $ model }

\begin{figure}
\includegraphics[width=\linewidth]{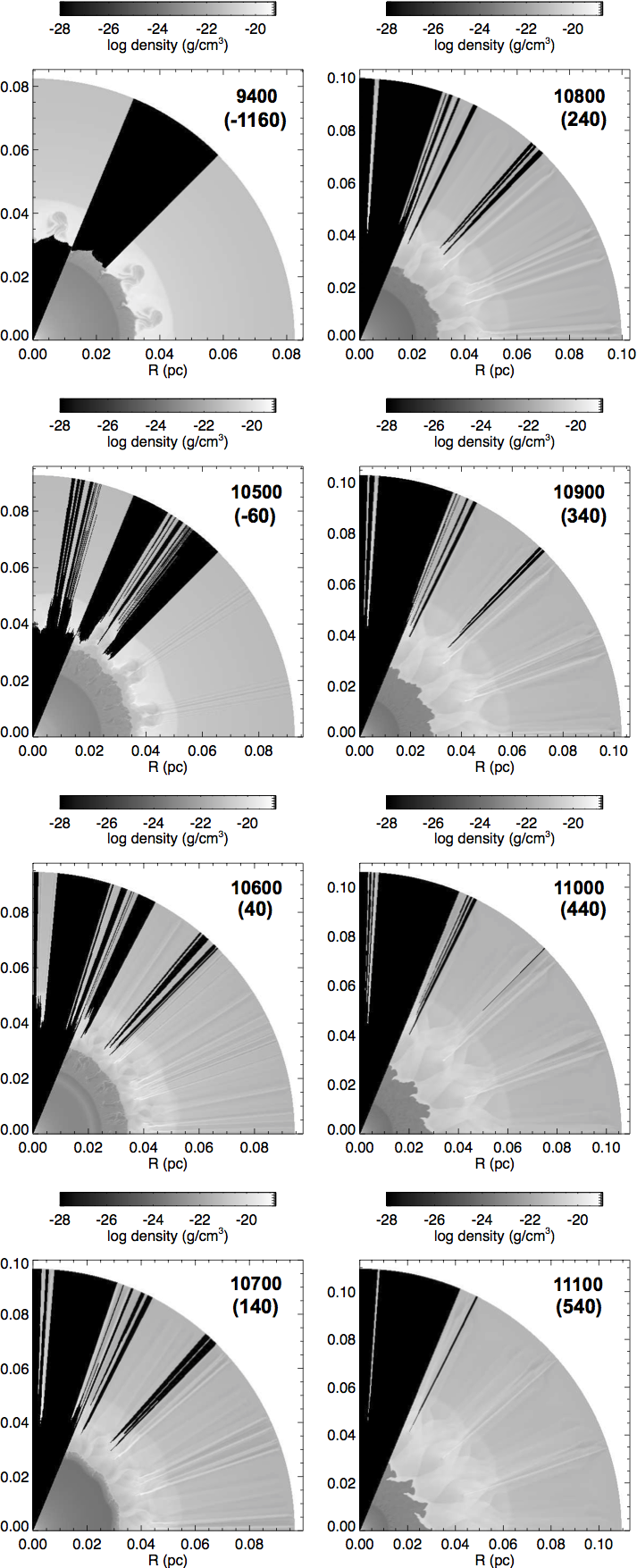}
\caption{Phase A for 2.0 \Mo}
\label{f9}
\end{figure}

\begin{figure}
\includegraphics[width=\linewidth]{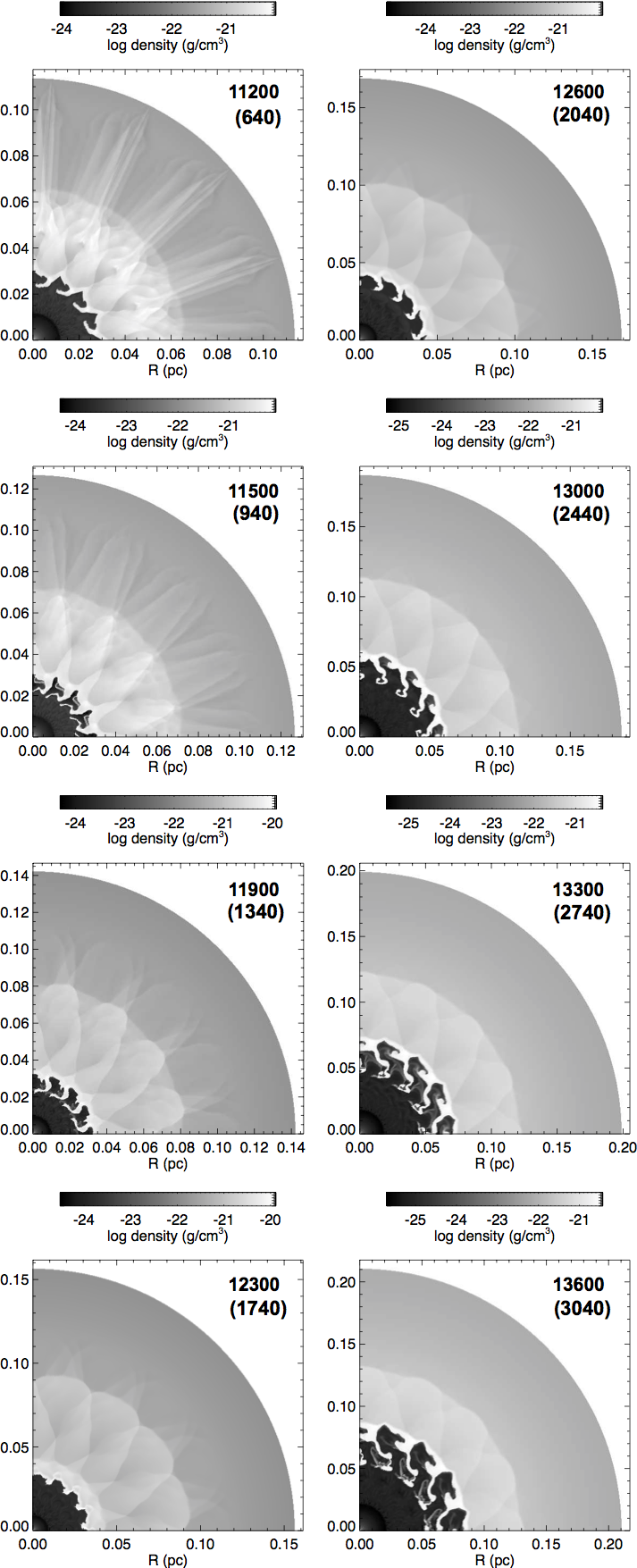}
\caption{Phase B for 2.0 \Mo}
\label{f10}
\end{figure}

\begin{figure}
\includegraphics[width=\linewidth]{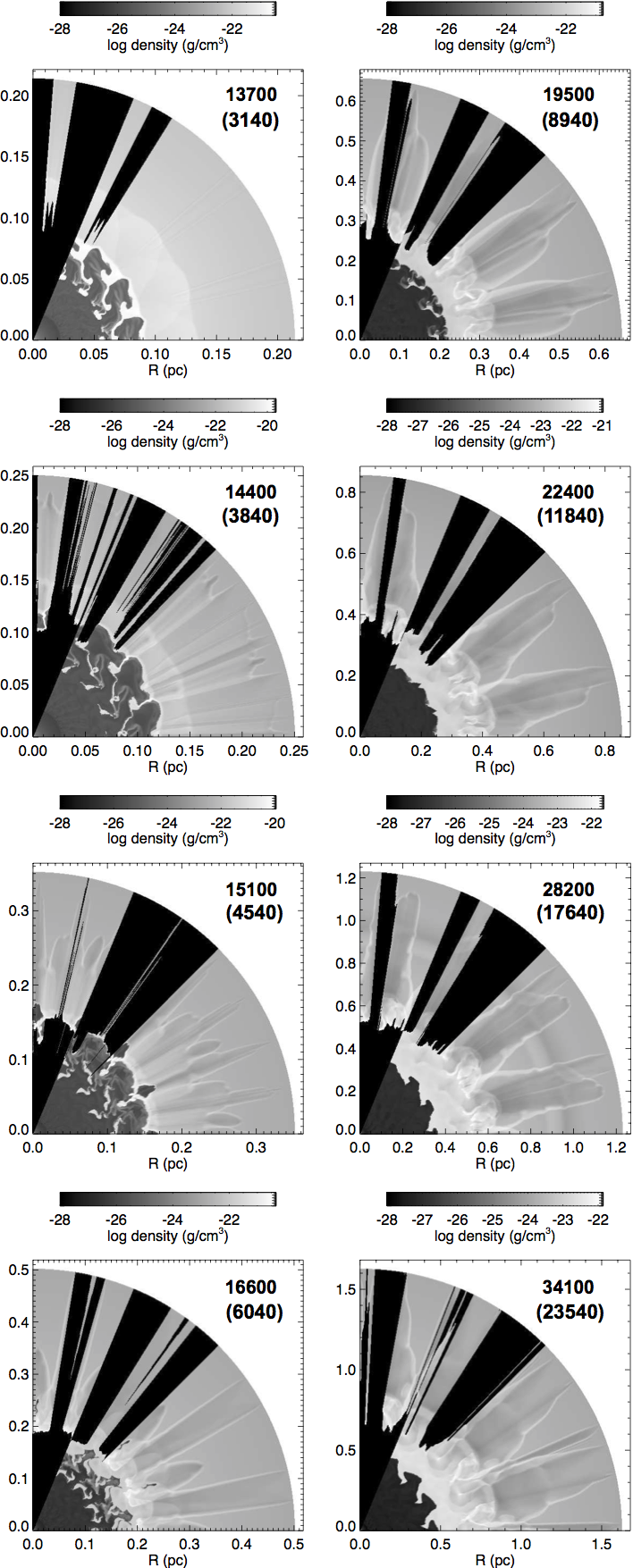}
\caption{Phase C for 2.0 \Mo}
\label{f11}
\end{figure}

Figures 9, 10, and 11 show the evolution of the $2 \, \Mo$ model.
In the first snapshot in Figure 9, we see a young planetary nebula
with an internal shock at 0.027 pc, a contact discontinuity
at 0.032 pc, and an external shock at 0.045 pc. As in the previous cases,
the dynamics is highly radiative in this phase, with shocked gas at
$10^5$ K. In the next snapshot in Figure 9, at 10,500 (-60) years,
the ionization front makes its way through the nebula, beginning
to create the first neutral spikes.
The spikes survive for 600 years until the end of phase A at
11,100 (540) years. Here, the bubble already contains hot gas at $10^6$ K, with an internal shock at 0.012 pc and a contact discontinuity at 0.03 pc.

In Figure 10, we see phase B. In the first snapshot at 11,200 (640) years, we see that the spikes have been photoionized, so it would be possible to observe these structures as light beams in $H_\alpha$.
This is actually observed in the nebula NGC 2867, as we will discuss in the discussion. Due to photoionization, these structures thermally expand and form many substructures in the halos, visible up to the last snapshot in Figure 10. At the end of phase B, at 13,300 (2,740) years, we have a nebula with an internal shock at 0.021 pc, a contact discontinuity at 0.082 pc, an external shock at 0.087 pc, and an attached halo up to 0.132 pc.

Figure 11 shows Phase C. This phase almost coincides with the entry into the cooling track (Figures 2 and 3), as can be seen in Table 1, where the maximum Teff of 172,981 K occurs at 14,887 (4,447) years. At this point, the ionizing photons have fallen from a maximum of 47.14 (in log) to 45.53, creating optimal conditions for recombination, leading to the formation of neutral spikes that will last until the end of our computation. The final structure computed at 34,100 (23,540) years produces a bubble with an internal shock at 0.05 pc, a contact discontinuity at 0.49 pc, a nearly extinct and fragmented swept-up shell with an external shock at 0.85 pc, and a halo alternating between neutral structures and $H_\alpha$ beams.

\subsection{The evolution of the  2.5 $ \Mo$ model}

\begin{figure}
\includegraphics[width=\linewidth]{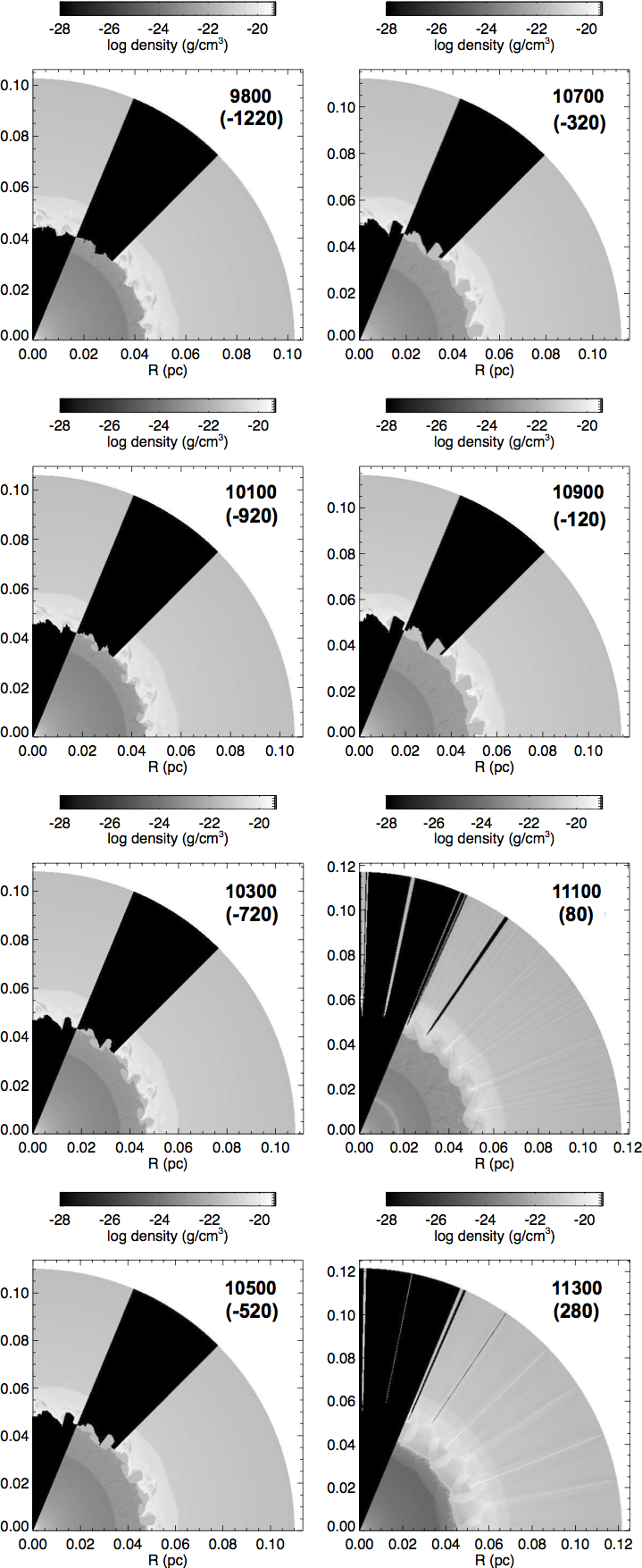}
\caption{Phase A for 2.5 \Mo}
\label{f12}
\end{figure}

\begin{figure}
\includegraphics[width=\linewidth]{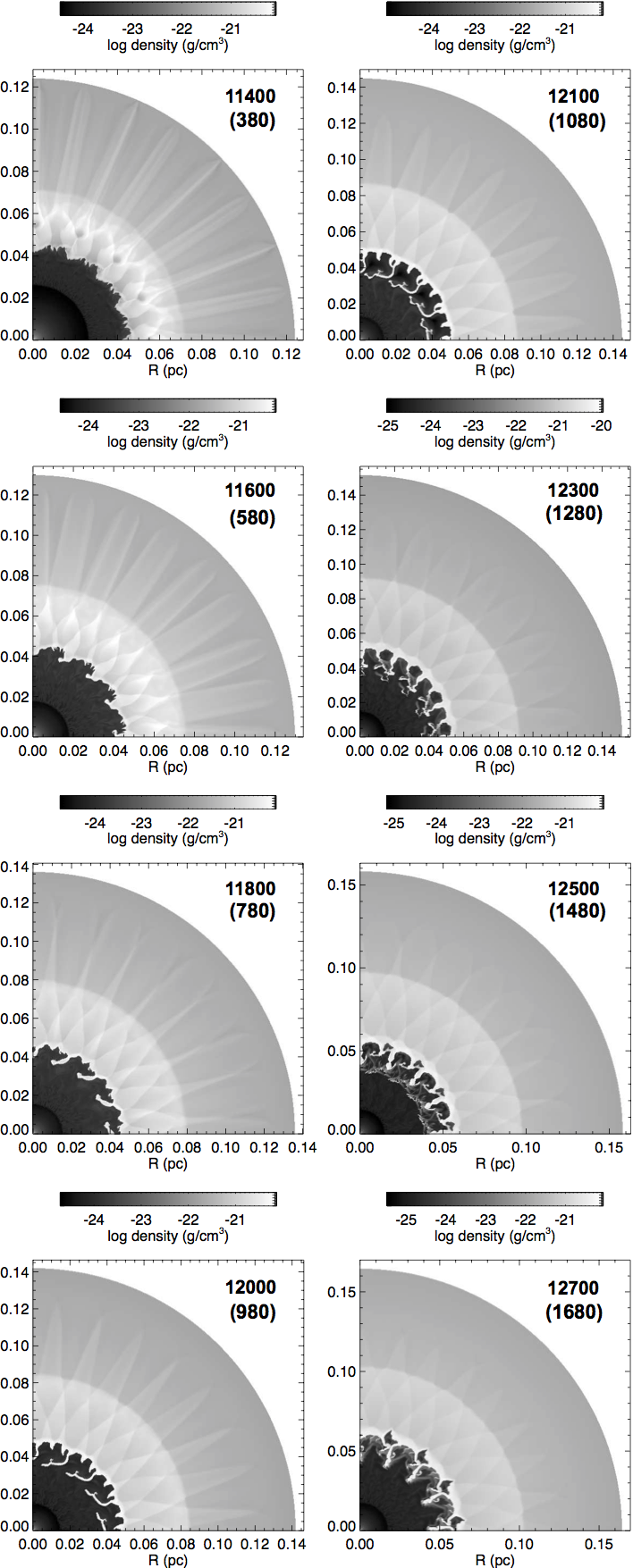}
\caption{Phase B for 2.5 \Mo}
\label{f13}
\end{figure}

\begin{figure}
\includegraphics[width=\linewidth]{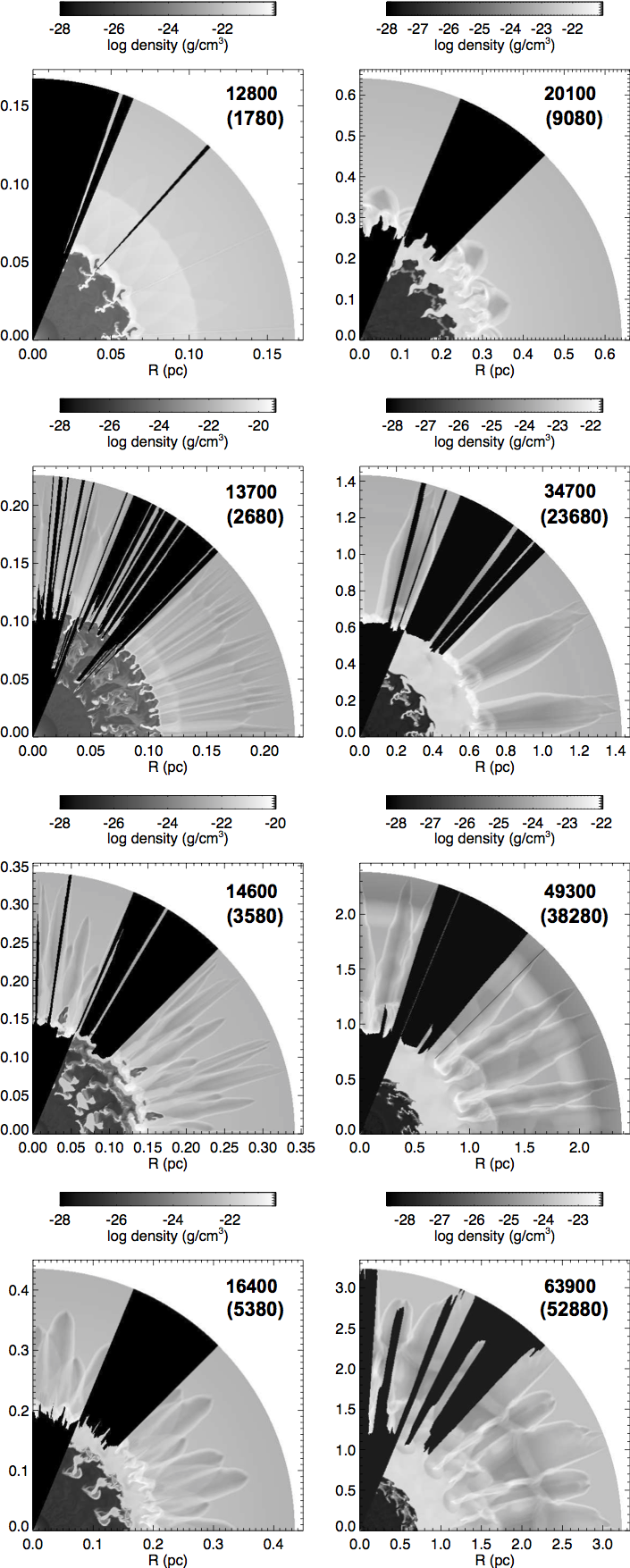}
\caption{Phase C for 2.5 \Mo}
\label{f14}
\end{figure}

Figures 12, 13, and 14 show the evolution of the $2.5 \, \Mo$ model. Starting in Figure 12, we have a young planetary nebula in the radiative phase with an internal shock located at 0.037 pc, a contact discontinuity at 0.044 pc, and an external shock at 0.057 pc. The transition from optically thick to optically thin occurs very rapidly in this model, as this central star evolves more rapidly than the previous ones. This transition occurs between the sixth and seventh snapshots after 10,900 (-120) years. It lasts less than 200 years, and the neutral spikes persist for only another 200 years, until phase A ends at 11,300 (280) years. Thus, the probability of observing these neutral spikes in this initial phase will be very small for this stellar model.

Phase B in Figure 13 initially shows an external halo with a lot of photoionized radial structure, which could be observed in $H_\alpha$ or [OIII], for example, and which gives way to a substructure in the attached halo that can also be observed in the optical range in several planetary nebulae, as we will see in the discussion section.
This substructure in the attached halo is observable during the whole phase B.

At the end of phase B at 12,700 (1,680) years, the bubble has an internal shock located at 0.0175 pc, a contact discontinuity at 0.06 pc, and an external shock at 0.065 pc. 

Figure 14 shows phase C, where recombination begins behind each dense clump formed in the swept-up shell, beginning at 12,800 (1,780) years. 

The wind reaches its maximum mechanical luminosity between snapshots 1 and 2, which is why a thickening of the swept-up shell is observed after snapshot 2, entering the cooling track at 2,666 years (Table 1) from photoionization. 

After 900 years in the second snapshot, we observe a large formation of neutral spikes, very similar to those observed in the Ring Nebula and Southern Ring Nebula in terms of their angular frequency, as we will see in the discussion section.
In the second snapshot, at 13,700 (2,680) years, we observe a large number of clumps within the bubble, immersed in its hot gas.
These clumps originate from the fragmentation of the 'trunks or horns' observed in the first snapshot in Figure 14 due to the drop of thermal pressure in the hot bubble.
We will discuss them in more detail in the discussion section.
The neutral spikes last until the fourth snapshot and almost completely disappear in the fifth snapshot at 20,100 (9,080) years.
Finally, starting with snapshot 6 at 34,700 (23,680) years, new neutral spikes form again as the
swept-up shell becomes optically thin in some areas, favoring their formation. By the end of our computation at 63,900 (52,880) years, we observe large neutral spikes embedded in a large \hii region, whose cometary heads are located around 1.3 pc in radius, and a hot bubble whose contact discontinuity is located at 0.6 pc.

\subsection{The evolution of the  3.5  $\Mo$ model }

\begin{figure}
\includegraphics[width=\linewidth]{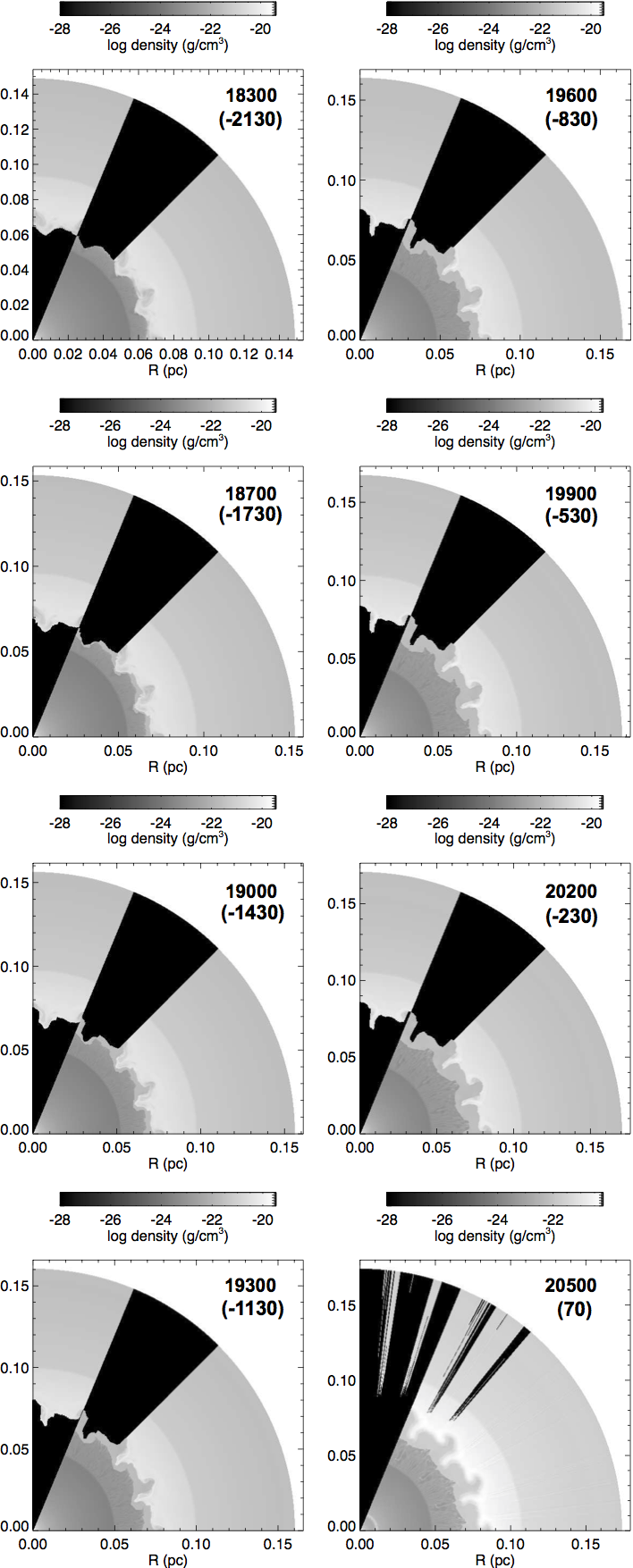}
\caption{Phase A for 3.5 \Mo}
\label{f15}
\end{figure}

\begin{figure}
\includegraphics[width=\linewidth]{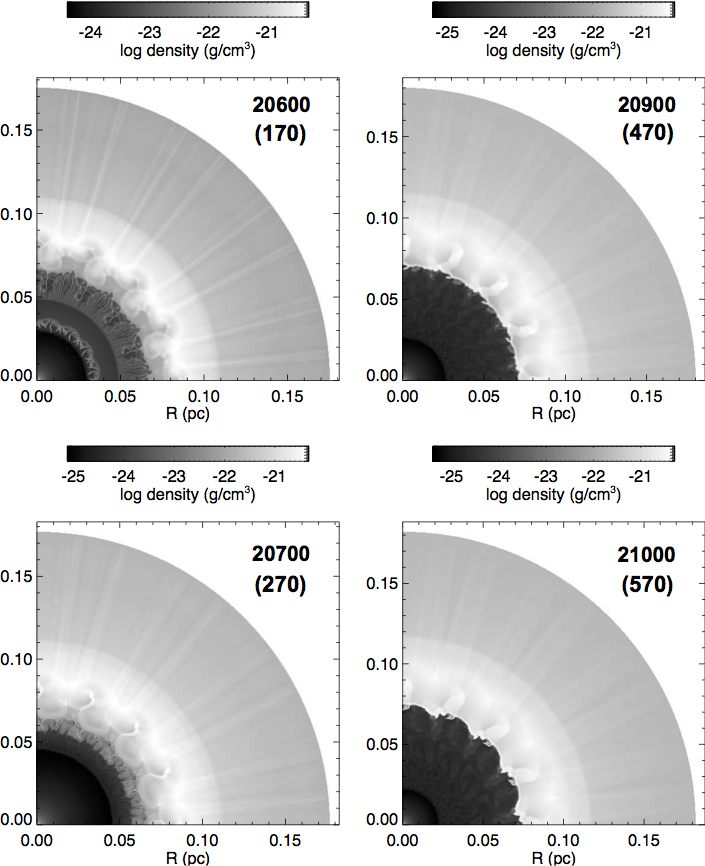}
\caption{Phase B for 3.5 \Mo}
\label{f16}
\end{figure}

\begin{figure}
\includegraphics[width=\linewidth]{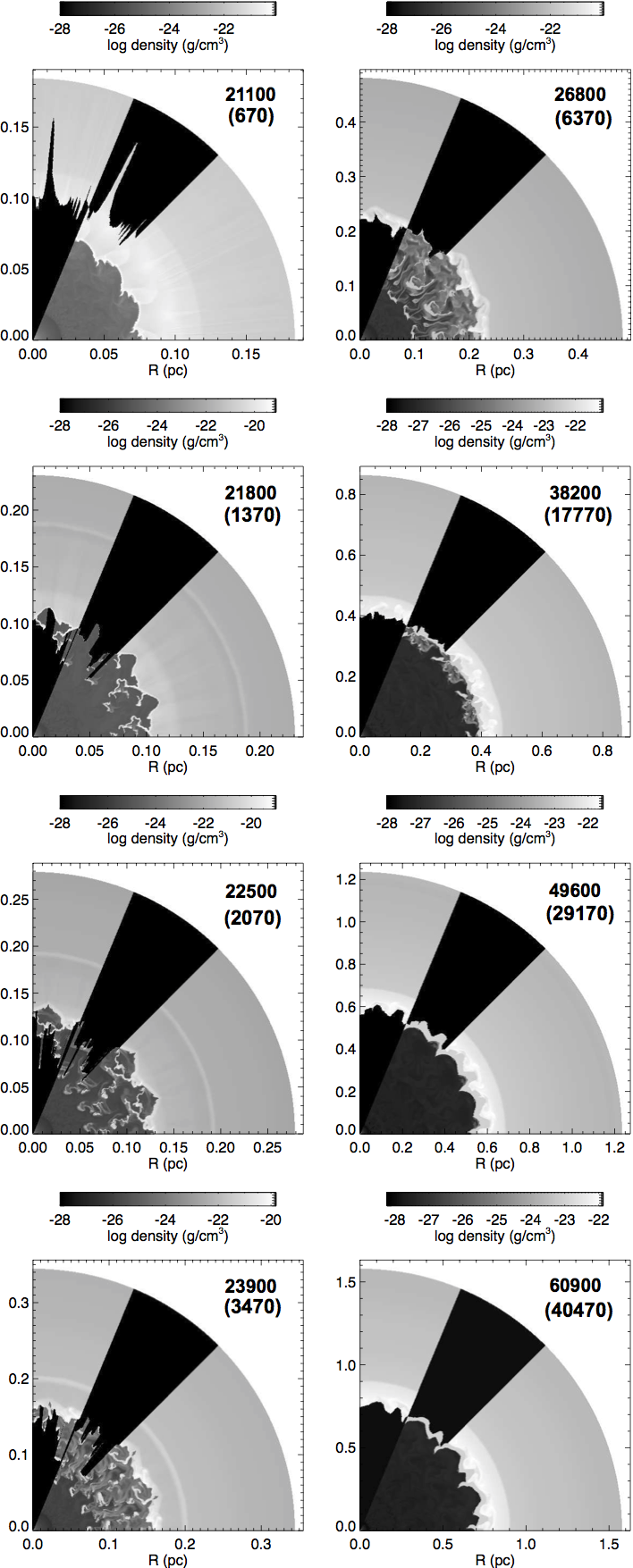}
\caption{Phase C for 3.5 \Mo}
\label{f17}
\end{figure}

Phase A of the $3.5 \, \Mo$ model is shown in Figure 15.
Initially, the planetary nebula has an internal shock located at 0.055 pc, a contact discontinuity at 0.065 pc, and an external shock at 0.094 pc.
The ionization front begins to cross the swept-up shell after 20,200 (-230) years, and the lifetime of the neutral spikes here is less than 300 years.

Figure 16 shows a very short phase B, only about 500 years long.
Although the first snapshot shows photoionized spikes, they are low-contrast ones because they did not have much time to form in phase A. Therefore, there is not much substructure in the outer part of the nebula, as was the case in lower mass models.

This stellar model evolves so rapidly that by the second snapshot, at 20,700 (270) years, it has already passed through the maximum number of ionizing photons, which occurs at 
20,629 (199) years  with $T_{eff}= 46,558$ K (Table 1).

In the fourth snapshot, at 21,000 (570) years, the inner shock is at 0.023 pc, the contact discontinuity at 0.074 pc, an outer shock at 0.0746 pc, and the exterior of the attached halo at 0.116 pc.

The transition to phase C is shown in Figure 17. Recombination occurs very rapidly, in less than 100 years, and does not allow time for neutral spikes to form. This model also forms a large number of clumps embedded in the hot bubble gas.

The maximum in mechanical luminosity occurs at 21,497 (1.067) years, between snapshots 1 and 2, where it begins to enter into the cooling track.

It can be seen that in the fourth snapshot at 23,900 (3,470) years, some of these clumps are still optically thick to ionizing radiation, but they cease to be so in the fifth snapshot at 26,800 (6,370) years, so they begin to evaporate and fade from this point on.
In the last snapshot at 60,900 (40,470) years, we observe a planetary nebula with a thin \hii region with a radius of 0.75 pc.
The internal shock is located at 0.05 pc, and the contact discontinuity is located at 0.7 pc.

In summary, this model does not produce any neutral spikes throughout its entire evolution that can be observed.

\subsection{The evolution of the  5 $\Mo$ model}

\begin{figure}
\includegraphics[width=\linewidth]{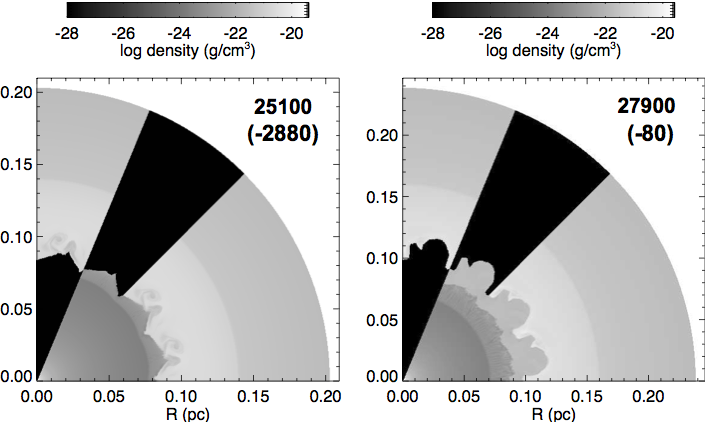}
\caption{Phase A for 5.0 \Mo}
\label{f18}
\end{figure}

\begin{figure}
\includegraphics[width=\linewidth]{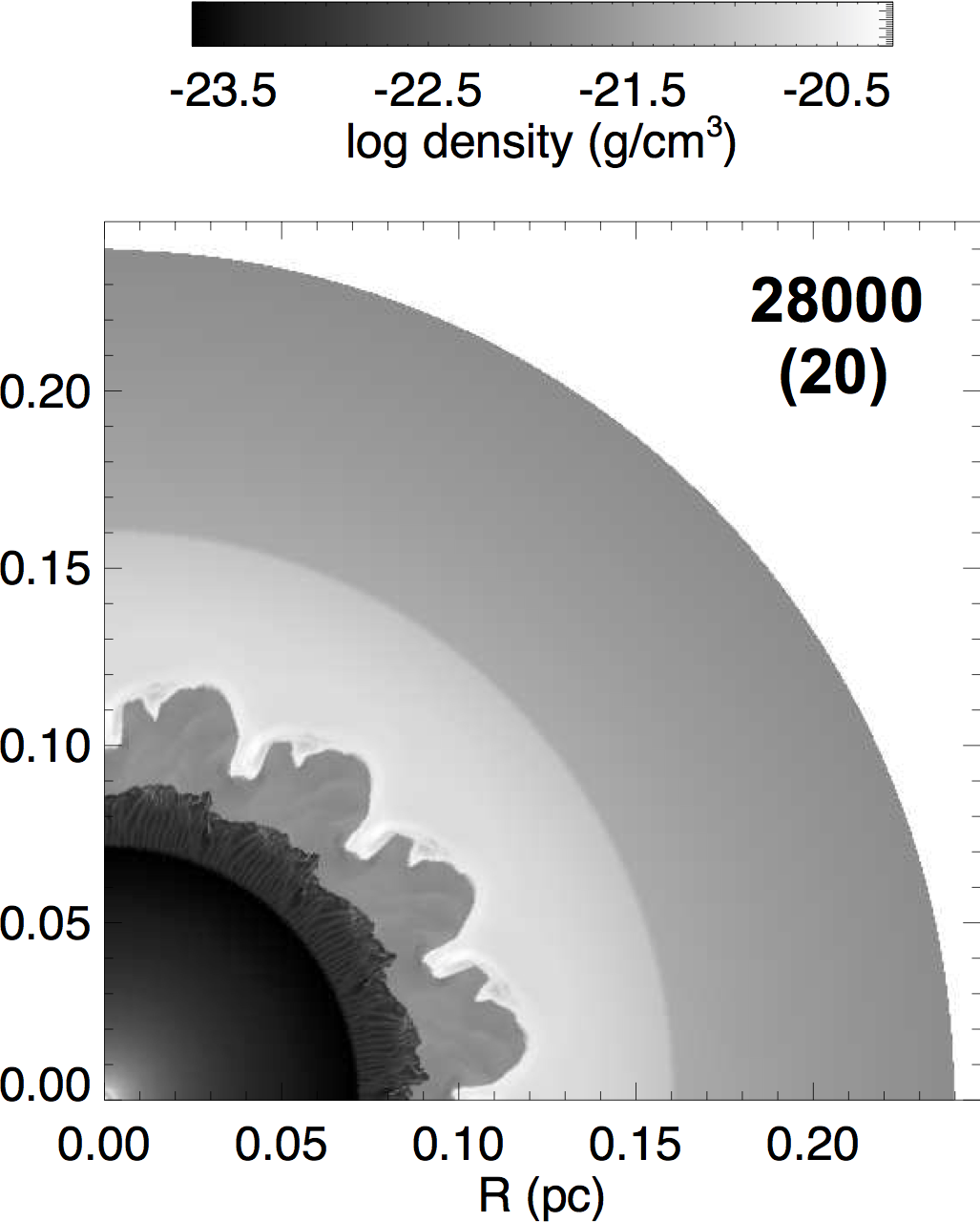}
\caption{Phase B for 5.0 \Mo}
\label{f19}
\end{figure}

\begin{figure}
\includegraphics[width=\linewidth]{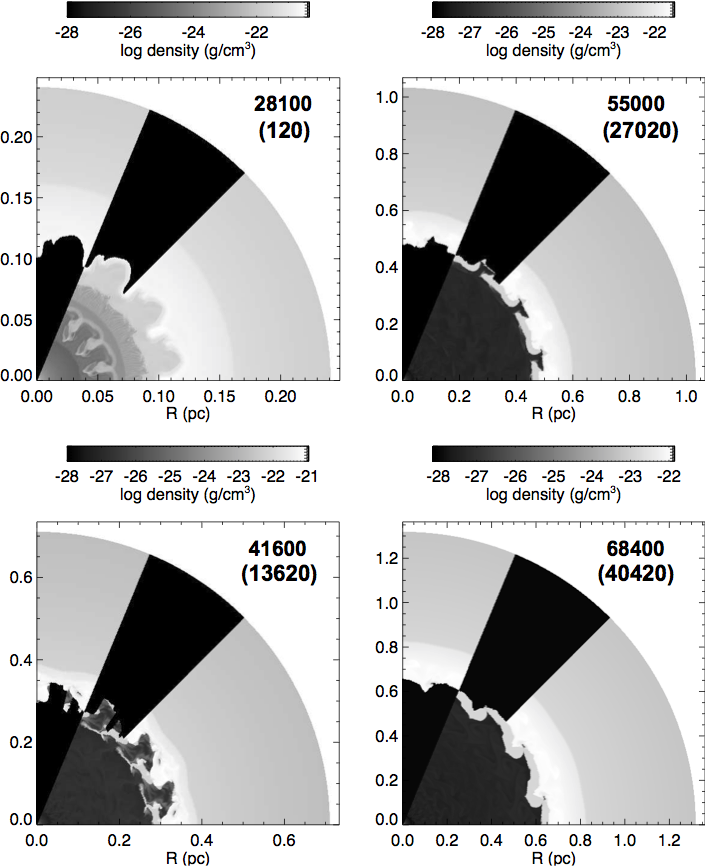}
\caption{Phase C for 5.0 \Mo}
\label{f20}
\end{figure}

Phase A of the $5.0 \, \Mo$ model is shown in Figure 18.
Initially, the planetary nebula has an internal shock located at 0.076 pc, a contact discontinuity at 0.09 pc, and an external shock at 0.14 pc. In this young phase, when the star reaches a temperature of 9,406 K or higher, the low-density portion inside the swept-up shell begins to be photoionized.

As the star approaches a temperature of 10,000 K, it very quickly enters phase B, shown in Figure 19, at 28,000 (20) years.
The evolution of this stellar model is so rapid once the star reaches 10,000 K that it must be difficult, statistically speaking, to observe a planetary nebula of this stellar mass in this phase. The maximum number of ionizing photons occurs at 28,010.3 (20.3) years, just after the snapshot in Figure 19. Phase B is very short, only about 100 years.

The transition to phase C is shown in Figure 20. Recombination occurs very rapidly, in less than 100 years, and does not allow time for neutral spikes to form. The maximum in mechanical luminosity occurs at 28,131.7 (141.7) years, just after the first snapshot in Figure 20, when it begins to enter the cooling track.
The first snapshot shows an internal swept-up shell  by the fast wind, which only lasts about 100 years. The mechanical wind injection in this model is so rapid and short in time that its dynamics resembles an explosion more than a constant wind injection.

In the last snapshot at 68,400 (40,420) years, we observe a planetary nebula with a thin \hii region with a radius of 0.66 pc.
The internal shock is located at 0.045 pc and the contact discontinuity at 0.625 pc.

As in the previous model, it can be said in summary that this model does not produce neutral spikes throughout its entire evolution that could be observed.

\section{Discussion}

\subsection{On the formation of clumps}

Based on our hydrodynamic models, we can confirm (see also Toal\'a \& Arthur 2014) that the clumps observed in many phases of 
the evolution are formed by thin-shell instabilities that rule the behavior of the swept-up shell by the stellar wind (Vishniac 1983; Garc\'{\i}a-Segura \& Mac Low 1995) or by the ionization front (Ionization-Shock Front instability; Garc\'{\i}a-Segura \& Franco 1996).
Clumps may or may not be optically thin to photoionizing radiation. This does not prevent them from changing from one state to another, depending on how ionizing radiation behaves during the evolution and the amount of material pilled-up by the instability.

The second snapshot in Figure 14 at 13,700 (2,680) years shows the formation of individual clumps due to the fragmentation of the ``trunks'' formed by the instability displayed in the first snapshot.
At this time, the central star has the following parameters: $\Mdot = 1.13 \times 10^{-10} \, \Moy$, $v_{\infty} = 14,183 \, \kms$, $T_{\rm eff} = 135,831$ K,
$F_{\star} = 1.62 \times 10^{45} \, s^{-1} $ (ionizing photons). The cooling track begins
at 13.646 (2,666) years, with a $T_{\rm eff} = 195,884 $ K and a 
$L_{\star} = 4.57 \times 10^{45} \,  s^{-1} $ of ionizing photons.

In Figure 21, we compare the clumps formed in the 2.5 $\Mo$ model with those observed, with the JWST, in the Ring Nebula, NGC 6720 (Wesson et al. 2024). The degree of similarity is quite striking, not only in terms of size but also in terms of spatial or angular frequency. It is interesting to note that the clumps in the simulations are embedded in a hot, shocked gas, so the surrounding gas moves subsonically. This produces vortices in the clumps called von Karman street vortices. This would explain why the clumps
seen in the James Webb Space Telescope observations look like winding snakes. Furthermore, it is observed that the shocked molecular hydrogen in the right part of Figure 21 surrounds the densest part of the clumps, observed in absorption in the left part of Figure 21. The explanation for this, based on the computed clumps shown in the center, is that the gas is being pulled away or eroded from the densest parts, which would be exposed to the ultraviolet radiation that produces fluorescence, but also could be heated up by a
conduction front.

\begin{figure}
\includegraphics[width=\linewidth]{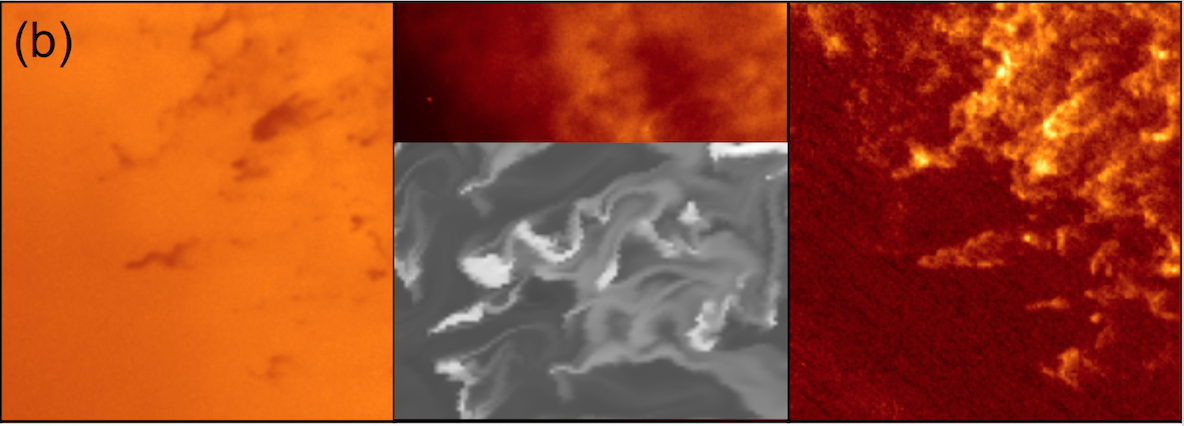}
\caption{Clumps in the Ring Nebula (Wesson et al. 2024) compared with modeled clumps in model 2.5 \Mo \, at 13,700 (2,680) years. }
\label{f21}
\end{figure}

Another point is that the clumps eventually expand, becoming completely photoionized and
thermally expanding into the hot bubble, finally getting diluted.
This is in part because they have become completely photoionized, but the real reason is that the thermal pressure of the hot gas has decreased considerably due to the decrease in the kinetic energy injected by stellar wind. The hot bubble depressurizes quite quickly once the star enters the
cooling track.
The clumps are in pressure equilibrium with their surroundings.
This pressure balance prevents the clumps from expanding while they are in their initial stages, partially  photoionized, because they are confined by the thermal pressure of the hot gas.
Eventually, the pressure in this hot gas decreases, allowing the globules to expand and complete
photoionization, which causes the lowering of their density.
The kinetic energy of the wind (due to the rate of mass loss) decreases on the way to the white dwarf.
Note that the kinetic energy of the wind is transformed directly into thermal energy (pressure) in the internal reverse shock.
The reverse shock position is at 0.02 pc. This position is defined where the ram pressure of the wind is equal to the pressure of the hot bubble.

In the final destruction of the clumps, not only the drop in thermal pressure plays an important role,
which causes the photoionization of the clumps and their expansion.
Photodissociation also plays an important role in the destruction of hydrogen molecules, since there are many photons between 913 \AA \, and 1,100 \AA, which are optically thin for neutral hydrogen.

In conclusion, once the clumps form, their lifetimes are dependent on the thermal pressure of the hot bubble, as well as on the combination of photoionization, heat conduction, and photodissotiation, the latter not included in the simulations.

\subsection{Subestructure in photoionzed haloes} 

\begin{figure}
\includegraphics[width=\linewidth]{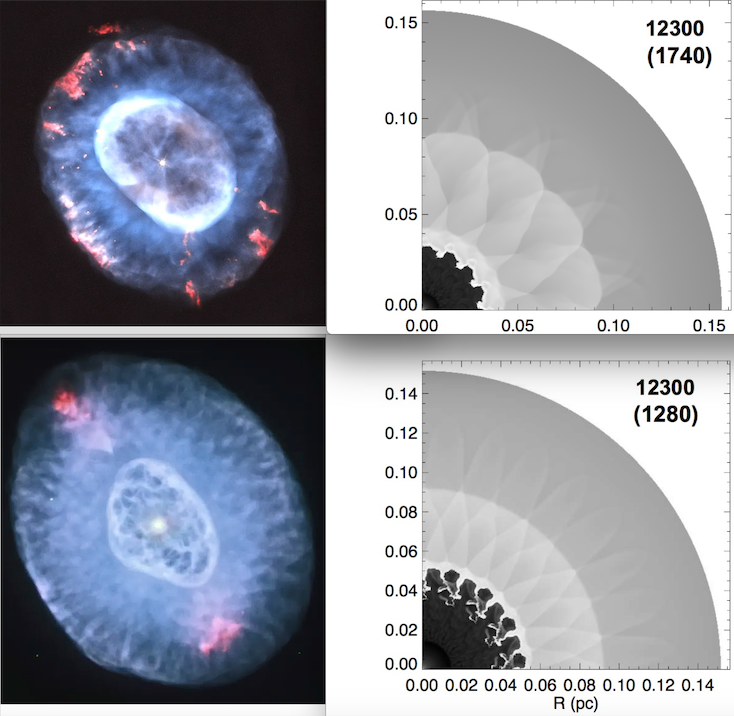}

\caption{Substructure in the haloes of NGC 7662 (up) and NGC 6826 (down) compared with models 2.0 \Mo \, and 2.5 \Mo. HST images are shown at highly altered contrast in order to emphasize their relevant features. Left: the spatial distribution of observed emission measures, $(n_e)^2 \times path length$, for NGC 7662 (upper left) and  6826 (lower left)  in the lines of [O III] (blue) and
[N II] (red).  Each color frame has been normalized to a peak value of unity.  An extended nebular halo has been subtracted in both cases.  Right: the best corresponding 2-D H+ density  distribution from the models. Credit: NASA/ESA Hubble Space Telescope.}
\label{f22}
\end{figure}

We have seen that neutral spikes form in phases A in the first four models computed. These spikes, when fully photoionized in phases B, expand thermally, forming a significant amount of substructures in the photoionized halos outside the swept-up shells. These substructures often take on a triangular shape as they expand, resulting in collisions between them.
In Figure 22, we show two specific cases: two snapshots of the $2.0$ and $2.5 \, \Mo$ models, which are compared with the nebulae NGC 7662 and NGC 6826, respectively.
These substructures can be clearly seen in the HST images, but are most evident in the NGC 6826 nebula, where the triangular shapes of these substructures are clearly visible, especially in the lower right corner. This mechanism for forming the substructures is quite natural as it arises solely from hydrodynamic history and does not require the assumption that there are previous clumps formed in the AGB wind. 
Note that our comparison of NGC 7662 and NGC 6826 with the models is only morphological, to show that these substructures in the haloes can form by the expansion of the spikes 
once they are ionized. We do not know the masses of the progenitor stars, so we cannot compare the models with these nebulae quantitatively.
For example, NGC 7662, at a distance of 1643 pc (Gonz\'alez-Santamar\'{\i}a et al. 2021), 
has diameters of $ 32 \times 26 $ arcsec, giving $ 0.25 \times 0.22 $  pc. With an
expansion velocity of 26 $\kms$ measured at [OIII] (Hippelein, Baessgen, \& Grewing 1985),
it has a kinematic age of 4704 yr.
In the first snapshot of the 2 \Mo $\,$ model in Figure 22, the diameter is only 0.064 pc,
the expansion velocity is 38 \kms, and this would give a kinematic age of only 824 yr. 
Note that the kinematic age of the model does not match the age of the model
(1740 yr since photoionization), since the former assumes a constant expansion velocity. 
This is not the case, as the expansion velocity increases until the wind decays (Figure 2).
For NGC 6826, with a distance of 1275 pc (Gonz\'alez-Santamar\'{\i}a et al. 2021),
an expansion velocity of 27 \kms in [OIII] (Chu et al. 1984),
and diameters of $27 \times 24$ arcsec give us $ 0.167 \times 0.148 $ pc, giving a kinematic age
of 3025 yr. In the second snapshot of the 2.5 \Mo $\,$  model in Figure 22,
the diameter is 0.094 pc, with an expansion velocity of 43 \kms, giving us a kinematic age of 1070 yr.

Note that in these spherical models, we do not attempt to resolve the elliptical morphologies and the fast, low ionization emission regions ``FLIERs'' (Balick et al. 1994) observed in NGC 6826, as we assume that these features formed in previous proto-planetary phases with a possible origin in common envelope evolution (Garc\'{\i}a-Segura et al. 2021, 2022).   However, the numerous low-ionization structures ``LISs'' (Gon\c{c}alves et al. 2001) observed in NGC 7662 could be explained
by the overdensities observed at the end of the haloes.

Note also that our models are two-dimensional, which gives us a smaller number of substructures. The reader must imagine the actual three-dimensional version of these structures, which are actually cones in many different angles, and project them onto the plane of the sky. Thus, the number of substructures increases considerably, as is the case of NGC 6826.

Another characteristic of some halos, such as NGC 2867 (Corradi et al. 2003), is the appearance of light beams in $H_\alpha$. These are explained in the context of these models, since in the optically thin parts of the swept-up shells, ionizing radiation penetrates into the halos, producing photoionization (Garc\'{\i}a-Segura et al. 1999). It is precisely between these light beams that neutral spikes form.

\subsection{Neutral spikes}

The neutral spikes observed in hydrogen molecular lines, as in the Ring Nebula (Wesson et al. 2024) and the Southern Ring Nebula (De Marco et al. 2023), only form in places where nebulae are optically thick to ionizing radiation, usually due to dense clumps or filaments. 
An example of the formation of neutral spikes is shown in Figure 23 where we compare a snapshot of model 2.5 \Mo $\,$ with the James Webb Space Telescope observations of the Ring nebula
(Wesson et al. 2024).
Note that our comparison of NGC 6720 with the 2.5 \Mo \, model in
Figure 23 is also only morphological, to show how the spikes are distributed.
We do not claim that this model fits observations since, as in the
previous cases, we do not know the mass of the progenitor star.
In the special case of NGC 6720, the distance is 783 pc (Gonz\'alez-Santamar\'{\i}a et al. 2021) 
and the expansion velocity
in [NII] is 45 \kms (Guerrero, Manchado, \& Chu 1997).
With diameters of $ 90 \times 60 $ arcsec (Guerrero, Manchado, \& Chu 1997),
this gives sizes of $ 0.34 \times  0.228 $ pc, giving a kinematic age of 3696 yr.
On the other hand, O'Dell et al. (2013) calculated a kinematic age of 4000 yr.
In the case of the 2.5 \Mo $\,$ model snapshot in Figure 23, the diameter is 0.22 pc,
with a velocity of 49 \kms, giving a kinematic age of only 2199 yr, while the
age since photoionization is 2680 yr.

There are two scenarios for this to occur: the first one when the transition from Phase A to Phase B occurs and the second when the transition from Phase B to Phase C occurs. In the first case, when the ionization front is trapped in a dense clump, a shadow is created behind it that allows the gas to remain in a neutral state, and the surroundings ionized gas of this shadow pushes the neutral gas, pilling it up and forming the spikes. In the second case, the ionization front is trapped by a dense clump, and the shadow behind it allows the gas to recombine to a neutral state, with a recombination time that depends primarily on the density. The two cases, although similar, are not the same, since in the first case we may have molecular material already existing in the slow wind of the AGB, while in the second case the molecular material will have to be formed anew.

\begin{table}
\caption{Window to detect neutral spikes.}
\begin{tabular}{lcc}
\hline
 $ Model $ &  Neutral spikes in phase A & Neutral spikes in phase C  \\
 $\,$ &    yr  &           yr      \\
\hline
 1.0 \Mo &  $\sim 3,000 $ &  0  \\
\hline
1.5 \Mo &  $\sim 800$  & Almost all   \\  
\hline
2.0 \Mo &  $\sim 600$  & All \\
\hline
2.5 \Mo &  $\sim 200 $  & All \\
\hline
3.5 \Mo & $> 200 $ &  0 \\
\hline
5.0 \Mo &  0 &  0 \\
\hline
\end{tabular}
\end{table}

In Table 2, we summarize the intervals in which neutral spikes can be detected. In the first case described above (Phase A to B), the time window decreases with the mass of the model, ranging from 3,000 years in the 1 \Mo \, case to 0 for 5 \Mo. This has to do with the speed of evolution in the HR diagram from post-AGB to the white dwarf (Figure 3). In the second case (Phase B to C), only the 1.5, 2.0, and 2.5 \Mo \, cases allow us to see the neutral spikes for most of the remaining time. This last result is very interesting, since simply observing the neutral spikes would give us information, although not very precise, about the mass of the stars.

These results only apply to idealized spherical models, and in the absence of stellar velocity relative to the interstellar medium.
In both the velocity models and the cases where common envelope evolution occurs, new models will have to be computed in the future.

The notion that the neutral spikes could be trailing low-ionization tails behind neutral cometary knots like those in the Helix Nebula is not applied here (see Mellema et al. 1998). That would imply that the gas has to travel a long distance in the slow wind of the AGB, faster than the nebula, as the spikes can be physically very long.

\begin{figure*}
\includegraphics[width=\linewidth]{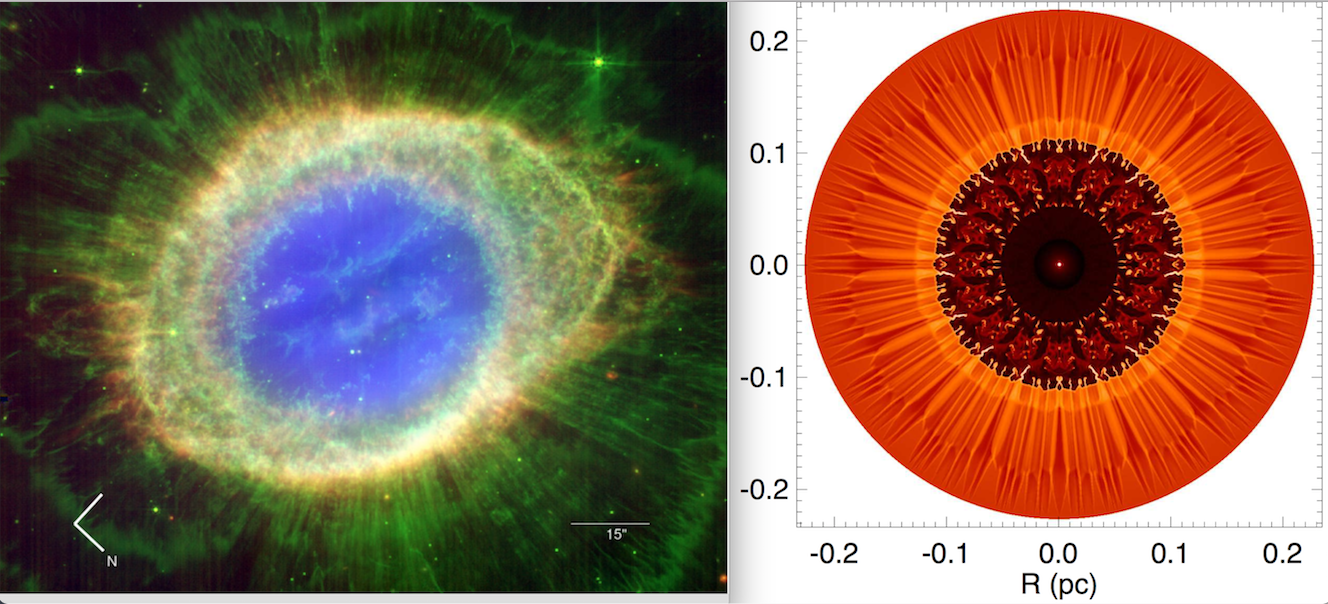}
\caption{The Ring Nebula seen in molecular Hydrogen compared with model 2.5 \Mo \, at 13,700 (2,680) years. Left: the spatial distribution of observed emission measures, $(n_e)^2 \times  path length$, for NGC 6720 in the lines of F2550W (blue), F560W (green), and F1130W (red).  Each color frame has been normalized to a peak value of unity (Wesson et al. 2024) (Credit: NASA/ESA/CSA James Webb Space Telescope). Right: the best corresponding 2-D total gas density  distribution of model 2.5 \Mo .} 
\label{f23}
\end{figure*}

\subsection{Final remarks}

We now consider the likely deviations from various idealized assumptions, initial conditions, and critical model evaluations. 

\begin{enumerate}

\item  Ejected mass and momentum of the slow winds.
The mass ejected in the AGB wind and its linear momentum depend on the mass-loss rate and its velocity.
To date, stellar evolution theory is somewhat imprecise regarding mass-loss rates in the
AGB phases and leads to considerable uncertainty. Similar formulas are used in stellar evolution models 
(Vassiliadis \& Wood 1993, Miller Bertolami 2016).
Furthermore, the growing evidence (De Marco et al. 2025)
that most nebulae arise from a common envelope evolution (CEE)
shows us that new calculations are needed in the future that take into account this violent expulsion of the AGB envelope, rather than the slow expulsion that is done in isolated stars. 
For example, the distribution of density, velocity, and asymmetries will be completely different in the cases of isolated stars and in the cases of CEE. In the former, all the imparted momentum will be determined by the momentum that stellar radiation can provide, while in the latter, the energy of the binary's orbit will be of vital importance. It is also very important to consider when the CEE occurs, 
whether at the beginning or end of the AGB, since this governs how long the star injected a slow AGB wind and how many thermal pulses were involved.
Regarding the sphericity of the winds, it is clear that CEE (Garc\'{\i}a-Segura et al. 2018 and references therein) completely break this view, as there is a tendency for the gas to be expelled toward the equator. 
In the calculations presented here, we have used the angular coordinate $\phi$,
which implies that we are using the equatorial plane, i.e., $\theta = 90^{\circ} $.
In the case of a CEE,
our present calculations are perhaps best applied to mid-latitudes around $45^{\circ}$,
since the CEE displaces a lot of gas toward the equator (higher density than these spherical models),
which would give optically thick situations throughout, and leaving the polar regions
devoid of gas (lower density than the spherical case), which would give optically thin situations from the start. 
Talking about CEE here does not mean that we can apply our models
to these scenarios, on the contrary, we believe it highlights our limitations with single-star
stellar models. In Garc\'{\i}a-Segura et al. (2014) and Garc\'{\i}a-Segura et al. (2016),
we proved that asymmetries can only be achieved with contact binaries,
and given that a large majority of planetary nebulae are not spherical, this suggests that 
new models that include binaries and CEE are absolutely necessary in the future.

\item  The initial geometry and temporal evolution of the fast winds.
When the initial distribution of the AGB gas is not spherical, as is the case in a CEE post-nebula, the evolution of the fast winds has an important impact on the fragmentation of PNe. These effects have been calculated by Garc\'{\i}a-Segura et al. (2022), where the fragmentation of the resulting nebula has an important dependence on the mass of the central star.
When the masses of the central stars are small, the fast winds take a long time to act, 
with ionizing radiation playing a much more important role in the evolution and fragmentation. For more massive stars, the effect is due to a combination of both simultaneously.
For practical purposes in this article, we have not been able to calculate aspherical
models, as that would require computation times of six months for each model. 
This is because to have the same resolution, but on grids that span 90 degrees at the 
$\theta$  angle, we would need to quadruple our angular resolution, i.e., $500 \times 800$ grids instead of $500 \times 200$ grids. This will be the subject of a future article.

\item  The molecular composition and geometry of the swept-up slow wind gas.
We are assuming that the AGB mass loss has spherical symmetry, as will
correspond to a single star or a wide binary with a separation of more
than 8 AU (Garc\'{\i}a-Segura et al. 2016), and has abundances of CNO equilibrium values.
However, the molecular content of AGB stars is very rich, with more than 90
molecules (Ag{\'u}ndez 2022) being the most abundant CO.
The last molecule, CO, is the most important in the cooling processes
according to Gail \& Sedlmayr (2013) in the AGB wind (see also
Berm\'udez-Bustamante et al. 2020).
As the temperature of the stars increases, most molecules are destroyed by
ultraviolet radiation. In the post-AGB phase, only 13 molecules remain (Ag{\'u}ndez
2022). When the star reaches 30,000 K only CO and $H_2$ are detected in PNe.
The $H_2$ is normally confined in the form of clumps and spikes (Manchado et al. 2015; De Marco
et al. 2022).
However, for bipolar PNe, these molecules are also observed in the equatorial regions
because of larger opacities.
We have not taken into account the effects of dust.
Dust production is very efficient in regions with high densities and temperatures below
the condensation value (~1,500 K). A secondary star in a binary system produces shocks in
the AGB wind in the form of spirals produced by the orbital motion. These shocks are fully
radiative, i.e., the post-shock temperature is not affected, but the jump in density is
proportional to the square of the Mach number.
Just a 10 km/s shock (i.e., similar to the orbital velocity) produces an increase of the
density by a factor of 100. Thus, the dust production should be much higher in the shocked
spirals of AGB winds than in AGB winds alone.
The other possibility is the formation of circumbinary rings, when the
mass ratio of the binary is above $q = M2/M1 > 0.78$ (Shu et al. 1979). These rings are also
formed by shocked AGB gas, either for detached or attached binary systems.
Thus, the higher density in these rings also contributes to a very effective dust formation.
Finally, the massive ejecta from a CEE event, at much lower velocities than a
supernova explosion, produce the best conditions for an effective production of dust.
Thus, what a binary system could produce is an ``amplification'' of dust production with
respect to a single AGB star in a wide range of scenarios, which are the subject of future
computations.

\item  Collimation and the azimuthal variation of obliquity of fast winds impact on the slower swept-up AGB winds.
It is true that clumps tend to be more common in elliptical and bipolar PNe.
Garc\'{\i}a-Segura et al. (2021) calculates how collimated jets or winds produce bipolar shapes in protoplanetary nebulae, which then produce elliptical or bipolar nebulae depending on the mass of the central star (Garc\'{\i}a-Segura et al. (2022).
When fast winds impact obliquely on the inner walls of nebulae, Kelvin-Helmholtz 
instabilities can occur. These instabilities also distribute the gas unevenly across latitudes, potentially producing regions that are more opaque than others. 
This may have an impact on the fragmentation produced by photoionization, as evidenced by Garc\'{\i}a-Segura et al. (2022). This apparently important effect will need to be addressed in a future article.

\item  The adopted cooling processes where the instabilities form clumps.
Cooling below 10,000 K is subject to many parameters and uncertainties, not only in the gas composition but also in the ionization fraction that it presents.
This zone below 10,000 K is crucial in clump formation, as it defines the temperature at which clumps form.
Very precise and computationally more expensive techniques (Berm\'udez-Bustamante et al. 2020) based on Gail \& Sedlmayr (2013) can be used for cases where computational and physical times are moderate. However, in cases where computational times are very long, the adopted cooling model needs to be simplified. Although this may not allow for a quantitative calculation, just as the numerical resolution does not, it is possible to obtain a first impression with more qualitative calculations. Therefore, to speed up the calculation, we opted for a linear interpolation in the cooling curve given by MacDonald \& Bailey (1981).
This does not imply that the calculated cooling is unrealistic. 
Comparisons with the two methods show small differences, with the effects of numerical resolution being more important than the cooling curve itself, since resolution is key in calculating the gas density.

\item  The cell size, grid boundary assumptions, and grid dynamic evolution in the model.
The resolution is the most important factor in any numerical simulation.
The higher the better, but this cannot be achieved in practice in many cases due to memory and computing time limitations.
Typically, to perform quantitative calculations for a single clump, we would need to have at least a $ 100 \times 100 $ region to resolve each clump. 
This, of course, can be done for the study of individual clumps, 
but it is not possible for us to do so in this context, where we are studying the formation of multiple clumps in a PN. High numerical resolution in a clump implies the resolution of its internal structure and density. Thus, with high resolution, higher densities can be achieved, which would make the clumps more opaque. Therefore, our calculations should be taken as a lower limit on the survival of a clump through the passage of the ionization front.
Our method of expanding the computational grid in time is the only way we currently have to achieve the maximum possible resolution, but it involves two additional problems. 
The first is that because of the external boundary inflow, the gas from the interstellar medium enters the grid as previously computed, and, in the case of the formation of long neutral spikes, their length will be truncated. However, here we are more interested in observing their existence rather than their actual length.
The second is the actual evolution of the expanding grid. This means that the resolution decreases over time. That is, clumps that form at late times have larger dimensions than those that form in the early stages. However, since thermal pressures are also lower at
late times, it is possible that this effect is partly real, i.e., 
the clumps of evolved PNs cannot be compressed as much, so their size would also be larger.

\item  The number of degrees of freedom in the model (2-D vs. 3-D).
When calculating the thin-shell instability in 3D, the clumps formed at the intersection of three bulges collect more gas than two bulges in 2D, due to the geometric effect.
This can be visualized as the foam produced by soapy water in a bathtub, where three small
bubbles join to form a common intersection point. In a shock with this characteristic, the swept-up gas is displaced to these points, forming a more massive and dense clump surrounded by three filaments, where each filament is the intersection of two bubbles.
Therefore, if it were possible to perform our calculations in 3D, we would possibly calculate longer survival times for neutral clumps before being photoionized in the early stages of evolution.
It is obvious that our calculations cannot currently be done in 3D, as they would require a computational time of 300 months per model, using a resolution of $500 \times 200 \times 200$, which is not
very large in terms of memory. 
However, we can perform the calculation in the future for a single clump and see in 3D and high resolution what its structure and survival would be. 
This is a good goal for a future article.

\item  The evolving extent to which the ionization front and its leading compression shock affect clump formation.
We have seen that thin-shell instability (Vishniac 1983) is a very logical mechanism for creating clumps due to the lateral displacement of the gas in the corrugations.
This instability, produced by the thermal pressure of the colliding wind, has its counterpart in ionization fronts, when they have a leading compression shock (D-type front) that is pushed by the thermal pressure of the \hii region.
This shock can also become corrugated and form clumps. This instability is called I-S front instability (Giuliani 1979; Garc\'{\i}a-Segura \& Franco 1996).
Thus, both winds and photoionization are important in the formation of clumps due to corrugated shocks.

\item   The effects of even minor amounts of turbulence in the early days of structural evolution.
Turbulence is always present in the ISM gas. Stellar winds also present some degree of
turbulence, including AGB slow winds. Although turbulence has probably a scale size that cannot
be resolved in the grids that we are using, it could have an important role triggering
thin shell instabilities in nature, since this instability evolves from short to long
wavelengths (Garc\'{\i}a-Segura \& Mac Low 1995). Thus, turbulence can produce earlier the
formation of instabilities.

\item    The use of other hydro codes and dynamic grid optimization using identical initial conditions.
The code and coordinate system to use for a specific problem is something that one must decide before tackling the problem. In general, our experience over the years indicates that for radial flows such as stellar winds and the formation of planetary nebulae, spherical coordinates are the most appropriate. Ideally, an AMR (adaptive mesh refinement) scheme would also be desirable in these coordinates.  Many codes that use AMR are optimized for Cartesian coordinates. While this can provide a high level of resolution in individual clumps, 
it can also be computationally expensive due to the small time steps resulting from the Courant condition. Another problem with Cartesian or cylindrical coordinates is that the initial region where the wind is defined is usually very small in size, 
so the initial evolution can be affected, which is not the case in spherical coordinates where the initial angular resolution is high. 
The homologous self-expansion scheme we use here is not trivial in codes that use Cartesian coordinates, allowing us to evolve from very small scales to several orders of magnitude larger. To do the same with a Cartesian code using AMR, one has to define a very large primary grid and use an AMR scheme with several orders of refinement to achieve a grid resolution of 5 or 6 orders of magnitude smaller in order to resolve the initial evolution. A new version of ZEUS-3D with AMR will soon be available in the public domain.

\item  The use of alternate but equally credible models for late- and post-AGB stellar evolution.
Stellar evolution codes use similar recipes for mass-loss rates, where the greatest source of uncertainty lies. The results are generally very similar for the evolution of the fast wind, since they use the studies by Pauldrach et al. (1988) and Bl\"ocker (1995). 
Regarding the AGB phase, studies by, for example, Vassiliadis \& Wood (1993) and Miller Bertolami (2016) uses formulas based on the pulsation periods of AGBs. 
In the post-AGB phase, Miller Bertolami's (2016) calculations, for example, give shorter transition times.
However, rather than focusing on a particular evolutionary model, 
we must keep in mind that none of them is based on a CEE, which is currently gaining popularity in the community. Therefore, the non-inclusion of a CEE would be the greatest source of uncertainty for all models. 
It is clear that a new generation of models that include CEE scenarios is of great importance for the field of PNs.

In short, all of these issues may affect the numbers, opacity, and growth/destruction 
trajectories of delicate clumps whose wind and radiation shadows are the fundamental reason for neutral spikes.

\item Differences in metal abundances from one PN to another might affect the number, densities, 
formation rates, and lifetimes of the clumps. It is logical to think that the more metals 
there are in the gas, the more efficient the cooling will be.
The more the gas cools, the thinner the swept-up shells can be, provided that photoionization 
does not dominate the thickness of the swept-up shell due to the thermal pressure of the 
ionized gas. Therefore, further cooling can, in principle, produce a larger number of 
clumps, since the thin-shell instability fragments the gas with a higher angular 
frequency. See for example the diferences in cut-off cooling temperatures of $10^4$ and $10^2$ K in Garc\'{\i}a-Segura and Mac Low (1995).
But this is something we can't easily verify in practice, not only because of the 
computation time each model takes (a month and a half), but also because we are 
primarily dominated by the effect of numerical resolution. So, to study the differences 
between one PN and another with different abundances, it is much more important to 
perform computations with much higher resolution in each clump first. 
This is impossible to do with our current numerical scheme.

\end{enumerate}

\section{Conclusions}

We can summarize the results of this article in the following three points.

\begin{itemize}
    
\item Clumps form at different stages of the planetary nebula evolution, always due to thin-shell instabilities, either due to winds (Thin-Shell instability; Vishniac 1983) or by the ionization front (Ionization-Shock Front instability; Garc\'{\i}a-Segura \& Franco 1996). Clumps are in pressure equilibrium with their surrounding medium and are subject to expansion or contraction depending on the pressure drops or rises of the surrounding medium, respectively, and their ionization status. We have found that when the flow is subsonic around the clumps, von Karma street vortices can form.

\item The substructure observed in many ionized halos is due to the expansion of previously formed neutral spikes. This substructure usually has conical shapes, which in projection appears to have triangular shapes.

\item Neutral spikes can be detected either at the formation of planetary nebulae or in their
late phases. In the first case, the temporal window decreases with the mass of the model,
ranging from 3,000 years in the 1 \Mo \, case to 0 for 5 \Mo. In the second case,
only the 1.5, 2.0, and 2.5 \Mo \, cases allow us to detect the neutral spikes for
most of the remaining time.

\item In these qualitative calculations, we are clearly limited by numerical resolution, 
since much higher resolution is required for each clump (at least 100 zones across) so that the calculation of density and internal structure provides a more precise and quantitative cooling.
This is something that must be addressed with a computational scheme more specific to this
problem.  Alternatively, simulations can be proposed with adaptive mesh refinement (AMR) codes 
or pseudo-lagrangian schemes in which the computational mesh moves in the clump reference plane, 
thus maintaining high spatial resolution regardless of spherical divergence in the expansion.
This will be the subject of a future article.

\end{itemize}

\section*{Acknowledgements}

We thank the referee for his/her comments that improved the manuscript considerably.
The authors thank Michael L. Norman and the Laboratory for Computational
Astrophysics for the use of ZEUS-3D. The computations
were performed at the Instituto de Astronom\'{\i}a-UNAM at  Ensenada.
GGS is partially supported by PAPIIT grant IG101223. GGS thanks UNAM and the DGAPA-PASPA grant for the sabbatical year at the IAA-CSIC. GGS also thanks IAA-CSIC and Estaci\'on Experimental IHSM La Mayora for his sabbatical stay.
AMT acknowledges the support from the State Research
Agency (AEI) of the Spanish Ministry of Science and Innovation (MCIN) under
grant PID2020- 115758GB-I00/AEI/10.13039/501100011033 and
the COST Action CA21126 - Carbon
molecular nanostructures in space (NanoSpace), supported by COST (European
Cooperation in Science and Technology).

\section*{Data availability}

The data underlying this article will be shared on reasonable request to the corresponding author.

%%%%%%%%%%%%%%%%%%%% REFERENCES %%%%%%%%%%%%%%%%%%

% The best way to enter references is to use BibTeX:

\bibliographystyle{mnras}
%\bibliography{example} % if your bibtex file is called example.bib

\begin{thebibliography}{99}

%\bibitem[\protect\citeauthoryear{Author}{2012}]{Author2012}
%Author A.~N., 2013, Journal of Improbable Astronomy, 1, 1

\bibitem[\protect\citeauthoryear{  }{  }]{  } Ag{\'u}ndez M., 2022, EPJWC, 265, 29

\bibitem[\protect\citeauthoryear{  }{  }]{  } Akras, S. Gonc\c{c}alves, D. R., Ramos-Larios, G., \& Aleman, I., 2020, Galaxies, 8, 30

\bibitem[\protect\citeauthoryear{  }{  }]{  } Balick, B., Perinotto, M., Maccioni, A., Terzian, Y., \& Hajian, A., 1994, \apj, 424, 800

\bibitem[\protect\citeauthoryear{  }{  }]{  } Berm\'udez-Bustamante, L. C., Garc\'{\i}a-Segura, G., Steffen, W., Sabin, L., 2020, \mnras, 493, 2606

\bibitem[\protect\citeauthoryear{  }{  }]{  } Bl\"ocker, T. 1995, A\&A, 297, 727

\bibitem[\protect\citeauthoryear{  }{  }]{  } Chu Y.-H., Kwitter K.~B., Kaler J.~B.,
Jacoby G.~H., 1984, PASP, 96, 598

\bibitem[\protect\citeauthoryear{  }{  }]{  } Clarke, D. A., 1996, \apj, 457, 291


\bibitem[\protect\citeauthoryear{  }{  }]{  } Corradi, R. L. M., Sch\"onberner, D.,  Steffen, M., \& Perinotto, M., 2003, \mnras, 340, 417

\bibitem[\protect\citeauthoryear{  }{  }]{  } De Marco, O., Akashi, M., Akras, S. et al., 2023, NatAs, 7, 234

\bibitem[\protect\citeauthoryear{  }{  }]{  } De Marco, O., Aleman, I., \& Akras, S. 2025, Encyclopedia of Astrophysics (edited by I. Mandel, section editor F.R.N. Schneider), Elsevier 

\bibitem[\protect\citeauthoryear{  }{  }]{  } Gail H.-P., \& Sedlmayr E., 2013, Physics and Chemistry of Circumstellar Dust Shells. Cambridge Univ. Press, Cambridge

\bibitem[\protect\citeauthoryear{  }{  }]{  } Garc\'{\i}a-Segura, G., \& Franco, J., 1996, \apj, 469, 171

\bibitem[\protect\citeauthoryear{  }{  }]{  } Garc\'{\i}a-Segura, G., Langer, N., R\'o\.zyczka, M., \& Franco, J., 1999, \apj, 517, 767 

\bibitem[\protect\citeauthoryear{  }{  }]{  } Garc\'{\i}a-Segura, G., \& Mac Low, M.-M., 1995, \apj, 455, 160

\bibitem[\protect\citeauthoryear{  }{  }]{  } Garc\'{\i}a-Segura, G., Ricker, P., \& Taam, R. E., 2018, \apj, 860, 19

\bibitem[\protect\citeauthoryear{  }{  }]{  } Garc\'{\i}a-Segura, G., Taam, R. E. , \& Ricker, P. M., 2021, \apj, 914, 111

\bibitem[\protect\citeauthoryear{  }{  }]{  } Garc\'{\i}a-Segura, G., Taam, R. E. , \& Ricker, P. M., 2022, \mnras, 517, 3822

\bibitem[\protect\citeauthoryear{  }{  }]{  } Garc\'{\i}a-Segura, G., Villaver, E., Langer, N., Yoon, S. -C., \& Manchado, A.  2014, \apj, 783, 74 

\bibitem[\protect\citeauthoryear{  }{  }]{  } Garc{\'\i}a-Segura G., Villaver E., Manchado A., Langer N., Yoon S.-C., 2016, \apj, 823, 142.

\bibitem[\protect\citeauthoryear{  }{  }]{  } Giuliani, J. L., 1979, \apj, 233, 280

\bibitem[\protect\citeauthoryear{  }{  }]{  } Gon\c{c}alves, D. R.,  Corradi, R. L. M.; Mampaso, A.  2001, ApJ, 547, 302

\bibitem[\protect\citeauthoryear{  }{  }]{  } Gonz\'alez-Santamar\'{\i}a I., Manteiga M.,
Manchado A., Ulla A., Dafonte C., L\'opez Varela P., 2021, A\&A, 656, 51

\bibitem[\protect\citeauthoryear{  }{  }]{  } Hora, J. L., Latter, W. B., Smith, H. A., \& Marengo, M., 2006, \apj, 652, 426

\bibitem[\protect\citeauthoryear{  }{  }]{  } Guerrero M.~A., Manchado A., Chu Y.-H., 1997, \apj, 487, 328

\bibitem[\protect\citeauthoryear{  }{  }]{  } Guerrero, M. A., Villaver, E., Manchado, A., Garc\'{\i}a-Lario, P., \& Prada, F., 2000, \apjs, 127, 125

\bibitem[\protect\citeauthoryear{  }{  }]{  } Hippelein H.~H., Baessgen M., Grewing M., 1985, 
A\&A, 152, 213

\bibitem[\protect\citeauthoryear{  }{  }]{  } Kastner, J. H., Gatley, I, Merrill, K. M., Probst, R., \& Weintraub, D., 1994, \apj, 421, 600


\bibitem[\protect\citeauthoryear{  }{  }]{  } Kwok, S., Purton, C. R., \& Fitzgerald, P. M., 1978, \apjl, 219, L125


\bibitem[\protect\citeauthoryear{  }{  }]{  } MacDonald, J. \& Bailey, M. E., 1981, \mnras, 197, 995


\bibitem[\protect\citeauthoryear{  }{  }]{  } Manchado A., Stanghellini L., Villaver E., Garc{\'\i}a-Segura G., Shaw R.~A., Garc{\'\i}a-Hern{\'a}ndez D.~A., 2015, \apj, 808, 115

\bibitem[\protect\citeauthoryear{  }{  }]{  } Mellema, G., Raga, A. C., Canto, J., Lundqvist, P., Balick, B., Steffen, W., Noriega-Crespo, A., 1998, A\&A, 331, 335

\bibitem[\protect\citeauthoryear{  }{  }]{  } Miller Bertolami, M. M. 2016, A\&A, 588, 25

\bibitem[\protect\citeauthoryear{  }{  }]{  } O'Dell C.~R., Ferland G.~J., Henney W.~J., Peimbert M., 2013, AJ, 145, 170

\bibitem[\protect\citeauthoryear{  }{  }]{  } Pauldrach, A., Puls, J., Kudritzki, R. H., M\'endez, R. H., \& Heap, S. R., 1988, A\&A, 207, 123

\bibitem[\protect\citeauthoryear{  }{  }]{  } Raymond, J. C. \& Smith, B. W., 1977, \apjs, 35, 419
             
\bibitem[\protect\citeauthoryear{  }{  }]{  } Shu, F.H., Lubow S.H., \& Anderson L., 1979, \apj, 229, 223

 \bibitem[\protect\citeauthoryear{  }{  }]{  } Stone, J. M. \& Norman, M. L., 1992, \apjs, 80, 753
             
\bibitem[\protect\citeauthoryear{  }{  }]{  } Toal\'a, J. A., \& Arthur, S. J., 2014, \mnras,  443, 3486

\bibitem[\protect\citeauthoryear{  }{  }]{  } Treffers, R. R., Chu, Y. -H., 1982, \apj, 254, 569

\bibitem[\protect\citeauthoryear{  }{  }]{  } Vassiliadis, E., \& Wood, P., 1993, \apj, 413, 641

\bibitem[\protect\citeauthoryear{  }{  }]{  } Vassiliadis, E., \& Wood, P., 1994, \apjs, 92, 125 

\bibitem[\protect\citeauthoryear{  }{  }]{  } Villaver, E., Manchado, A., \& Garc\'{\i}a-Segura, G., 2002, \apj, 581, 1204


\bibitem[\protect\citeauthoryear{  }{  }]{  } Vishniac, E. T., 1983, \apj, 274, 152

\bibitem[\protect\citeauthoryear{  }{  }]{  }  Wesson, R., Matsuura, M., Zijlstra, A. A. et al., 2024, \mnras, 528, 3392

\end{thebibliography}

% Alternatively you could enter them by hand, like this:
% This method is tedious and prone to error if you have lots of references

% Don't change these lines
\bsp	% typesetting comment
\label{lastpage}
\end{document}